\documentclass[aps,twocolumn,prb,showpacs,floatfix,superscriptaddress,citeautoscript]{revtex4-1}

\usepackage{epsfig,amsmath,amssymb}
\usepackage{graphics} 
\usepackage{subfigure}
\usepackage{color}
\usepackage{multirow}
\usepackage{mathrsfs}
\usepackage{natbib}

\begin{document}

\title
{
Frustrated spin-$\frac{1}{2}$ $J_{1}$--$J_{2}$ isotropic $XY$ model on the honeycomb lattice
}
\author
{R.~F.~Bishop}
\affiliation
{School of Physics and Astronomy, Schuster Building, The University of Manchester, Manchester, M13 9PL, UK}

\author
{P.~H.~Y.~Li}
\affiliation
{School of Physics and Astronomy, Schuster Building, The University of Manchester, Manchester, M13 9PL, UK}

\author
{C.~E.~Campbell}
\affiliation
{School of Physics and Astronomy, University of Minnesota, 116 Church Street SE, Minneapolis, Minnesota 55455, USA}

\begin{abstract}
  We study the zero-temperature ground-state (GS) phase diagram of a
  spin-half $J_{1}$--$J_{2}$ $XY$ model on the honeycomb lattice with
  nearest-neighbor exchange coupling $J_{1}>0$ and frustrating
  next-nearest-neighbor exchange coupling $J_{2} \equiv \kappa
  J_{1}>0$, where both bonds are of the isotropic $XY$ type, using the
  coupled cluster method.  Results are presented for the GS energy per
  spin, magnetic order parameter, and staggered dimer valence-bond
  crystalline (SDVBC) susceptibility, for values of the frustration
  parameter in the range $0 \leq \kappa \leq 1$.  In this range we
  find phases exhibiting, respectively, N\'{e}el $xy$ planar [N(p)],
  N\'{e}el $z$-aligned [N($z$)], SDVBC, and N\'{e}el-II $xy$ planar
  [N-II(p)] orderings.  The N\'{e}el-II states, which break the
  lattice rotational symmetry, are ones in which the spins of
  nearest-neighbor pairs along one of the three equivalent honeycomb
  directions are parallel, while those in the other two directions are
  antiparallel.  The N(p) state, which is stable for the classical version of the model in the range $0 \leq
  \kappa \leq \frac{1}{6}$, is found to form the GS phase out to a
  first quantum critical point at $\kappa_{c_{1}} = 0.216(5)$, beyond
  which the stable GS phase has N($z$) order over the range
  $\kappa_{c_{1}} < \kappa < \kappa_{c_{2}}=0.355(5)$.  For values
  $\kappa > \kappa_{c_{2}}$ we find a strong competition to form the
  GS phase between states with N-II(p) and SDVBC forms of order.  Our
  best estimate, however, is that the stable GS phase over the range
  $\kappa_{c_{2}} < \kappa < \kappa_{c_{3}} \approx 0.52(3)$ is a mixed state with both SDVBC and N-II(p) forms of order; and for values $\kappa > \kappa_{c_{3}}$ is the N-II(p)
  state, which is stable at the classical
  level only at the highly degenerate point $\kappa=\frac{1}{2}$.
  Over the range $0 \leq \kappa \leq 1$ we find no evidence for any of
  the spiral phases that are present classically for all values $\kappa > \frac{1}{6}$, nor for any quantum
  spin-liquid state.
\end{abstract}

\pacs{75.10.Jm, 05.30.Rt, 75.30.Kz, 75.40.Cx}

\maketitle

\section{INTRODUCTION}
\label{introd_sec}
In recent years many theoretical studies have been devoted to various
frustrated quantum spin models on the two-dimensional (2D) honeycomb
lattice, using several different quantum many-body
techniques \cite{Rastelli:1979_honey,Fouet:2001_honey,Mulder:2010_honey,
  Cabra:2011_honey,Ganesh:2011_honey,Clark:2011_honey,DJJF:2011_honeycomb,Albuquerque:2011_honey,
  Mosadeq:2011_honey,Oitmaa:2011_honey,Mezzacapo:2012_honey,PHYLi:2012_honeycomb_J1neg,Bishop:2012_honey_phase,Bishop:2012_honeyJ1-J2,RFB:2013_hcomb_SDVBC}.
Particular attention has focused on the spin-$\frac{1}{2}$
$J_{1}$--$J_{2}$ model in which nearest-neighbor (NN) pairs of spins
interact via an isotropic Heisenberg interaction with exchange
coupling parameter $J_{1}$, and next-nearest-neighbor (NNN) pairs
interact via a similar isotropic Heisenberg interaction with exchange
coupling parameter $J_{2}$.  When the NN interaction is
antiferromagnetic in nature (i.e., $J_{1}>0$), a corresponding
antiferromagnetic NNN interaction (i.e., $J_{2}>0$) acts to frustrate
the N\'{e}el order that is preferred by the NN bonds acting by
themselves.  The extended spin-$\frac{1}{2}$ $J_{1}$--$J_{2}$--$J_{3}$
model, in which the additional next-next-nearest-neighbor (NNNN)
Heisenberg bonds of exchange coupling strength $J_{3}$ are included,
has also been studied on the honeycomb lattice.  Both
spin-$\frac{1}{2}$ models have rich ground-state (GS) zero-temperature
($T=0$) phase diagrams and, indeed the nature of the ground state is
still not fully resolved beyond all doubt in various parts of the
phase diagrams.  The $J_{1}$--$J_{2}$--$J_{3}$ model in particular even
exhibits a diverse array of ordered GS phases in the classical case,
which corresponds to the limit $s \rightarrow \infty$ of the spin
quantum number $s$ of the spins residing on the honeycomb lattice
sites.

In view of the uncertainty that sill remains over the $T=0$ GS phase
diagram of the spin-$\frac{1}{2}$ $J_{1}$--$J_{2}$ Heisenberg model on
the honeycomb lattice, it is of great interest to examine closely
related models.  One such model is the isotropic frustrated
$J_{1}$--$J_{2}$ $XY$ model on the same honeycomb lattice.  Whereas
the isotropic $J_{1}$--$J_{2}$ Heisenberg model has the Hamiltonian,
\begin{equation}
{\cal H}_{{\rm H}}=J_{1}\sum_{\langle i,j \rangle}\mathbf{s}_{i}\cdot\mathbf{s}_{j}+J_{2}\sum_{\langle\langle i,k \rangle\rangle}\mathbf{s}_{i}\cdot\mathbf{s}_{k}\,,   \label{H_H}
\end{equation}
where index $i$ runs over all honeycomb lattice sites, indices $j$ and
$k$ run respectively over all NN and NNN sites to $i$, counting each
bond once only, and $\mathbf{s}_{i}=(s^{x}_{i},s^{y}_{i},s^{z}_{i}$)
is the spin operator (corresponding to a spin quantum number $s$) on
site $i$, the corresponding isotropic $J_{1}$--$J_{2}$ $XY$ model has
the Hamiltonian,
\begin{equation}
{\cal H}_{{XY}}=J_{1}\sum_{\langle i,j \rangle}(s^{x}_{i}s^{x}_{j}+s^{y}_{i}s^{y}_{j})+J_{2}\sum_{\langle\langle i,k \rangle\rangle}(s^{x}_{i}s^{x}_{k}+s^{y}_{i}s^{y}_{k})\,.  \label{H_XY}
\end{equation}

The two models of Eqs.\ (\ref{H_H}) and (\ref{H_XY}) on the honeycomb
lattice share exactly the same $T=0$ GS phase diagram in the classical
($s \rightarrow \infty$) limit
\cite{Rastelli:1979_honey,Fouet:2001_honey}, which makes it
particularly interesting to compare their corresponding phase diagrams
for the extreme quantum limiting case, $s=\frac{1}{2}$.  Even at the
classical level the two models on the honeycomb lattice share the
interesting feature that for values of the frustration parameter
$J_{2}/J_{1}\equiv\kappa > \frac{1}{6}$ the GS phase has an infinite
degeneracy
\cite{Rastelli:1979_honey,Fouet:2001_honey,Mulder:2010_honey}, since
in this regime the wave vector $\mathbf{Q}$ of the non-collinear
spiral states that form the stable GS phase can take any value over a
specific closed contour in the Brillouin zone, as we discuss more
fully in Sec.\ \ref{model_sec}.

One knows that novel quantum phases often emerge from such classical
models that exhibit an infinitely degenerate family of GS phases in
some region of phase space.  Typically one then finds that quantum
fluctuations lift this (accidental) GS degeneracy, either wholly or
partially, by the {\it order by disorder}
mechanism \cite{Villain:1977_ordByDisord,Villain:1980_ordByDisord}, to
favor either just one or several members of the classical family as
the quantum GS phase.  Indeed, for the present models it has been shown
\cite{Mulder:2010_honey} that spin-wave fluctuations at leading order,
$O(1/s)$, lift the accidental degeneracy in favor of specific wave
vectors, thus leading to spiral order by disorder.  Nevertheless, it
is well known that quantum fluctuations have the general tendency to
favor collinear ordering over noncollinear ordering, and one may thus
anticipate that for the spin-$\frac{1}{2}$ models in particular the
quantum fluctuations present should actually melt the spiral order for a wide
range of values of the frustration parameter, $\kappa$.  One such
collinear state, which is among the infinitely degenerate family of
ground states at the classical critical point $\kappa=\frac{1}{2}$, is
the so-called N\'{e}el-II state, in which all NN bonds along one of
the three equivalent honeycomb directions are ferromagnetic (i.e.,
with spins parallel), while those along the other two directions are
antiferromagnetic (i.e., with spins antiparallel).  This is precisely
the state favored by the above-mentioned spiral order by disorder
mechanism at the classical critical point $\kappa=\frac{1}{2}$.

Naturally, quantum fluctuations in the extreme $s=\frac{1}{2}$ quantum
limit can also be expected to destroy completely the magnetic order in
any (collinear or non-collinear) quasiclassical state in various
regions of the $T=0$ GS phase space, i.e., for various regions of the
frustration parameter $\kappa$.  Indeed, such magnetically disordered
regions have been observed in a large number of theoretical
calculations of the spin-$\frac{1}{2}$ isotropic $J_{1}$--$J_{2}$
Heisenberg model on the honeycomb lattice, which involve valence-bond
crystalline (VBC) phases with either plaquette or dimer ordering.

For example, in a recent calculation \cite{RFB:2013_hcomb_SDVBC} by
the present authors, using the coupled cluster method (CCM) carried
out to very high orders of approximation, the $T=0$ GS phase diagram
of the spin-$\frac{1}{2}$ isotropic $J_{1}$--$J_{2}$ Heisenberg model
on a honeycomb lattice was studied for the case $J_{1}>0$ in the range
$0 \leq \kappa \leq 1$ for the frustration parameter.  Four phases
were found, which exhibited, respectively, N\'{e}el, 6-spin plaquette,
staggered dimer, and N\'{e}el-II orderings, with corresponding quantum
critical points (QCPs) at $\kappa_{c_{1}} = 0.207 \pm 0.003$,
$\kappa_{c_{2}} = 0.385 \pm 0.010$, and $\kappa_{c_{3}} \approx
0.65 \pm 0.05$.  The two transitions at $\kappa_{c_{1}}$ [between
states with N\'{e}el and plaquette valence-bond crystalline (PVBC)
order] and $\kappa_{c_{3}}$ [between states with staggered dimer
valence-bond crystalline (SDVBC) and N\'{e}el-II order] were found to
be most likely of continuous second-order (and hence deconfined) type, while that
at $\kappa_{c_{2}}$ (between the two VBC states) was found to be most
likely of direct first-order type.  The competition between SDVBC and
N\'{e}el-II orderings was found to be particularly finely balanced,
and consequently the QCP at $\kappa_{c_{3}}$ has the largest
associated uncertainty.  Although broadly similar results have been
reported using other theoretical method,
\cite{Fouet:2001_honey,Mulder:2010_honey,
  Cabra:2011_honey,Ganesh:2011_honey,Clark:2011_honey,DJJF:2011_honeycomb,Albuquerque:2011_honey,
  Mosadeq:2011_honey,Oitmaa:2011_honey,Mezzacapo:2012_honey,PHYLi:2012_honeycomb_J1neg,Bishop:2012_honey_phase,Bishop:2012_honeyJ1-J2}
differences still remain and a complete consensus has not yet been
reached for the full $T=0$ GS phase diagram for the spin-$\frac{1}{2}$
Hamiltonian of Eq.\ (\ref{H_H}) on the infinite honeycomb lattice.

Nevertheless, since the CCM has proven itself to give very accurate
results for the GS phases and the associated QCPs for a very wide
range of other spin-lattice models on 2D lattices (see, e.g.,
Refs.\ \onlinecite{Bishop:1991_XXZ_PRB44,Zeng:1998_SqLatt_TrianLatt,Kruger:2000_JJprime,Bishop:2000_XXZ,Fa:2001_trian-kagome,Darradi:2005_Shastry-Sutherland,Schm:2006_stackSqLatt,Bi:2008_PRB_J1xxzJ2xxz,Bi:2008_JPCM_J1J1primeJ2,Darradi:2008_J1J2mod,Bi:2009_SqTriangle,Richter2010:J1J2mod_FM,Bishop:2010_UJack,Bishop:2010_KagomeSq,Reuther:2011_J1J2J3mod,DJJF:2011_honeycomb,Gotze:2011_kagome,Bishop:2012_checkerboard,Li:2012_honey_full,Li:2012_anisotropic_kagomeSq,Li:2013_chevron,Bishop:2013_crossStripe},
and references cited therein), it now seems an opportune time to apply
it also to the closely related spin-$\frac{1}{2}$ isotropic $XY$ model,
whose Hamiltonian is given by Eq.\ (\ref{H_XY}), in order to compare
its $T=0$ GS phase diagram with our previous results
\cite{RFB:2013_hcomb_SDVBC} for the spin-$\frac{1}{2}$ Heisenberg
model of Eq.\ (\ref{H_H}).  This has become particularly timely in
view of recent intriguing results \cite{Zhu:2013_honey_XY} for the
isotropic $XY$ model, using the density matrix renormalization group
(DMRG), which showed a stable GS phase in a relatively narrow region
of the frustration parameter $\kappa$, immediately beyond the QCP
below which N\'{e}el antiferromagnetic (AFM) collinear magnetic order
occurs in the $xy$ plane, in which N\'{e}el AFM order now occurs with
the spins aligned along the $z$ axis.  This is particularly surprising
in view of the total absence in the $XY$ model of any Ising-like terms
involving $s^{z}_{k}s^{z}_{l}$ between spins on any pairs of sites $k$
and $l$.  The spin-$\frac{1}{2}$ isotropic $XY$ model on the honeycomb
lattice had also been studied earlier \cite{Varney:2011_honey_XY} in
the context of an equivalent model of spinless hard-core bosons on the
honeycomb lattice at half-filling with NN and NNN hopping terms and
zero off-site interactions, using the exact diagonalization (ED) of
relatively small lattice clusters.  This work suggested the existence of
a particular quantum spin-liquid (QSL) GS phase in a similar regime of
phase space for the parameter $\kappa$.

One motivation for the later DMRG study \cite{Zhu:2013_honey_XY} was to
examine much larger finite lattice clusters than are feasible for ED
studies, in order to investigate whether the QSL GS phase might have
been an artefact of the small clusters studied, which would not
survive in the thermodynamic limit, $N \rightarrow \infty$, where $N$
is the number of lattice sites.  Indeed, the DMRG study
\cite{Zhu:2013_honey_XY} found no evidence for any QSL phases in the
$XY$ model.  By contrast, another very recent variational Monte (VMC)
study \cite{Carrasquilla:2013_honey_XY} of the isotropic $XY$ model,
which employed a variational QSL wave function of a particular type
(based on a decomposition of the bosonic particles into pairs of
spin-$\frac{1}{2}$ fermions, and including a long-range Jastrow factor
plus a Gutzwiller projection to enforce single bosonic occupancy)
found that such QSL states are energetically favored in the
intermediate frustration regime.

\begin{figure*}[!tbh]
\begin{center}
\mbox{
\subfigure[]{\scalebox{0.28}{\includegraphics{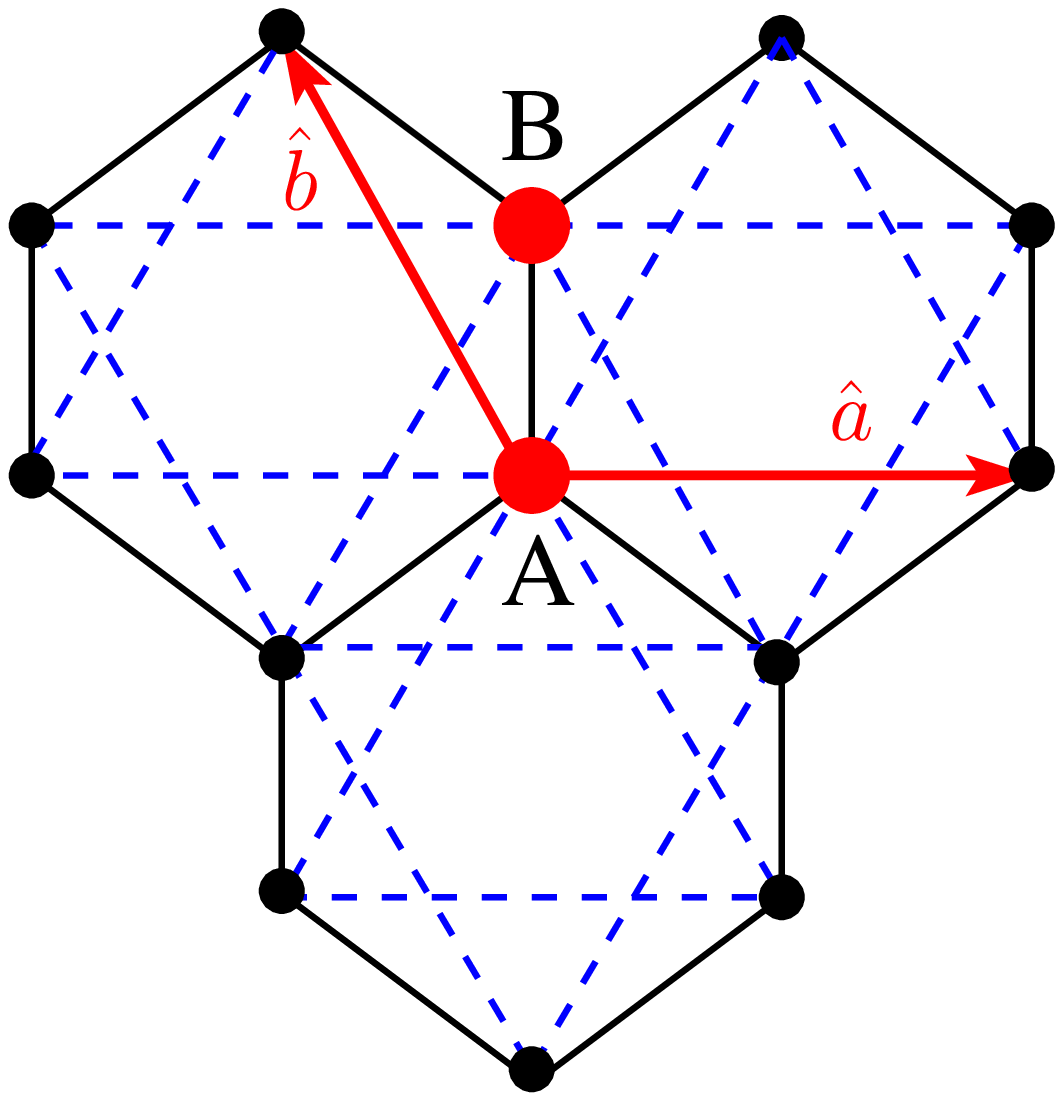}}}
\quad
\subfigure[]{\scalebox{0.2}{\includegraphics{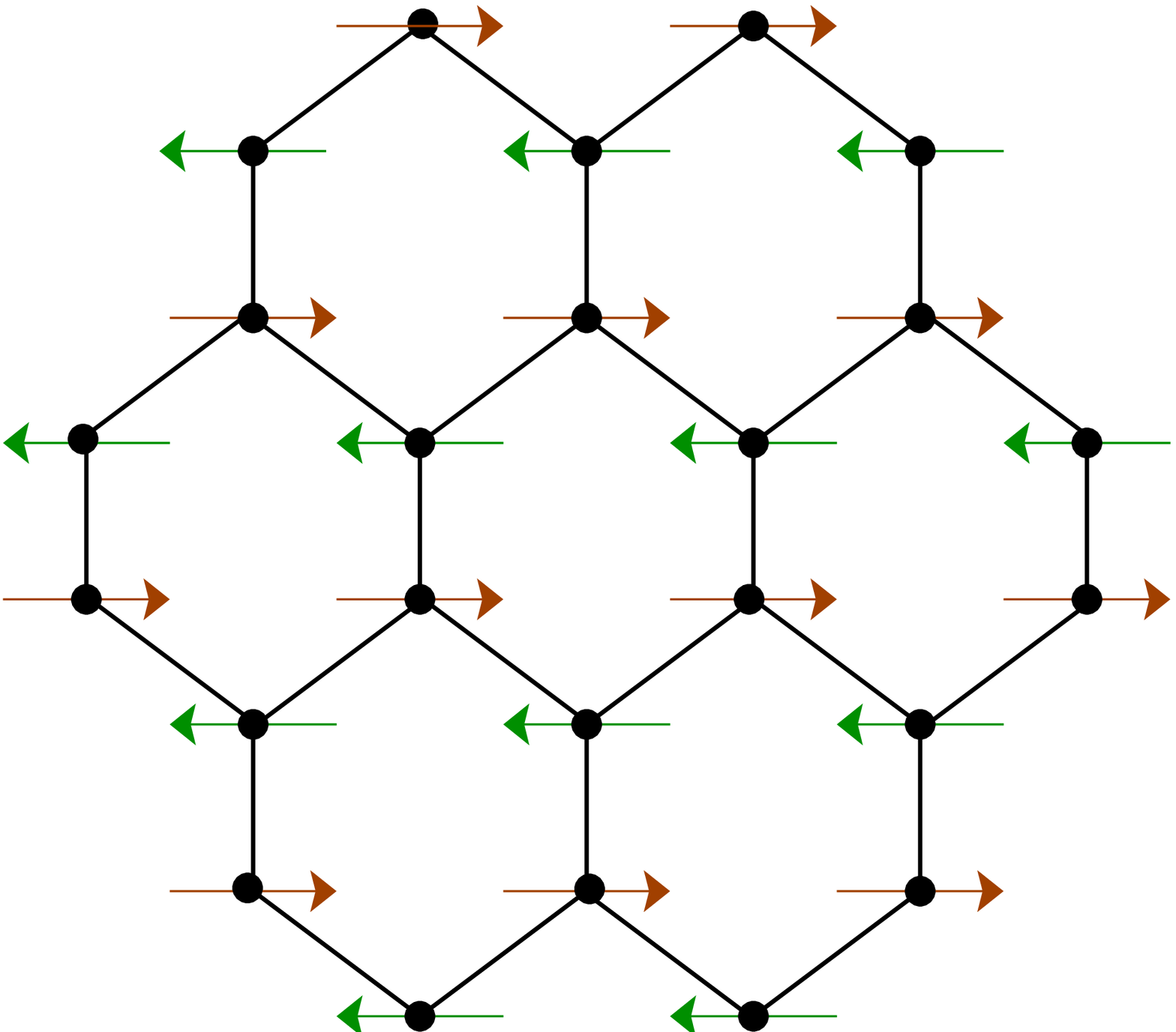}}}
\quad
\subfigure[]{\scalebox{0.2}{\includegraphics{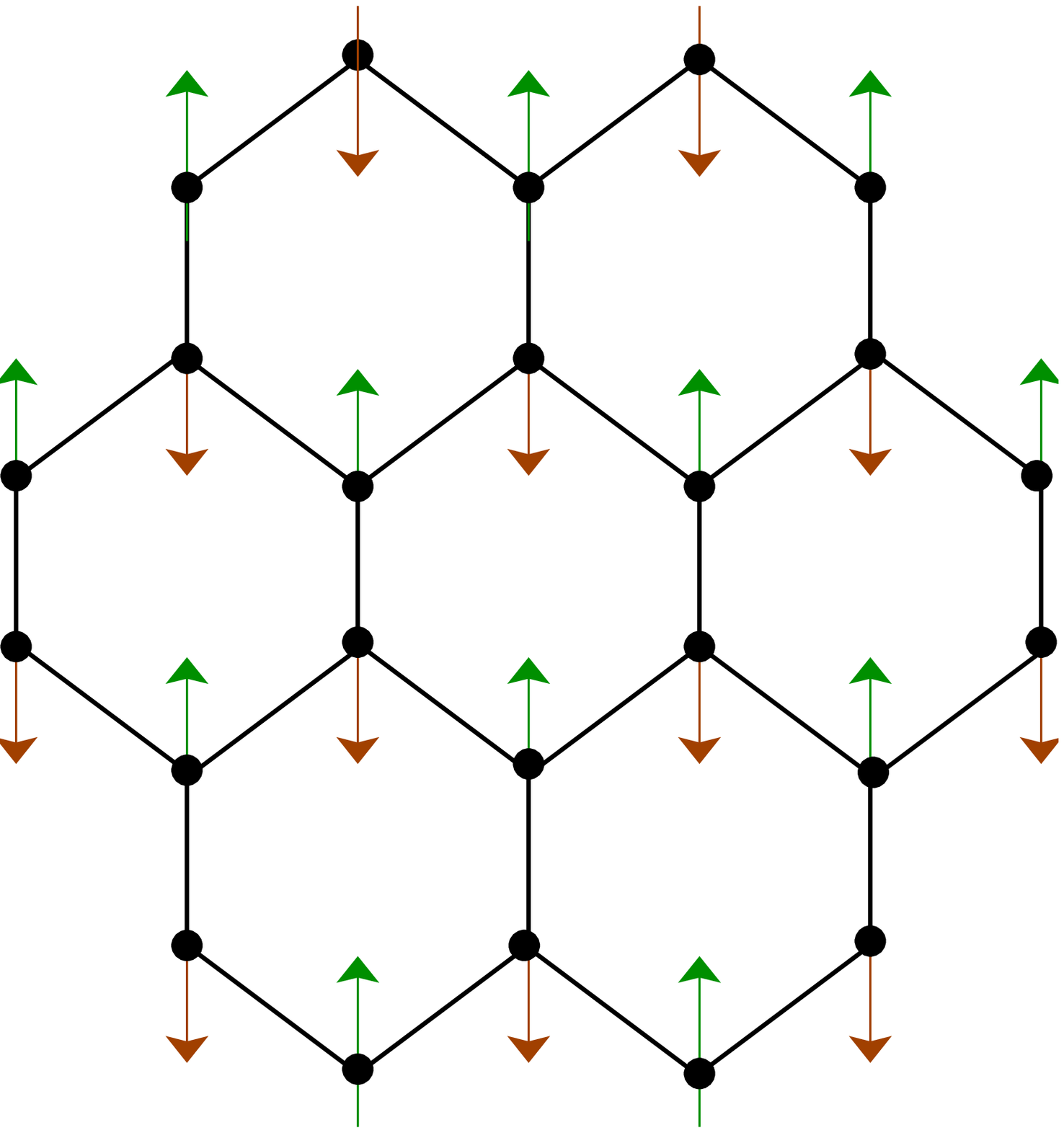}}}
\quad
\subfigure[]{\scalebox{0.2}{\includegraphics{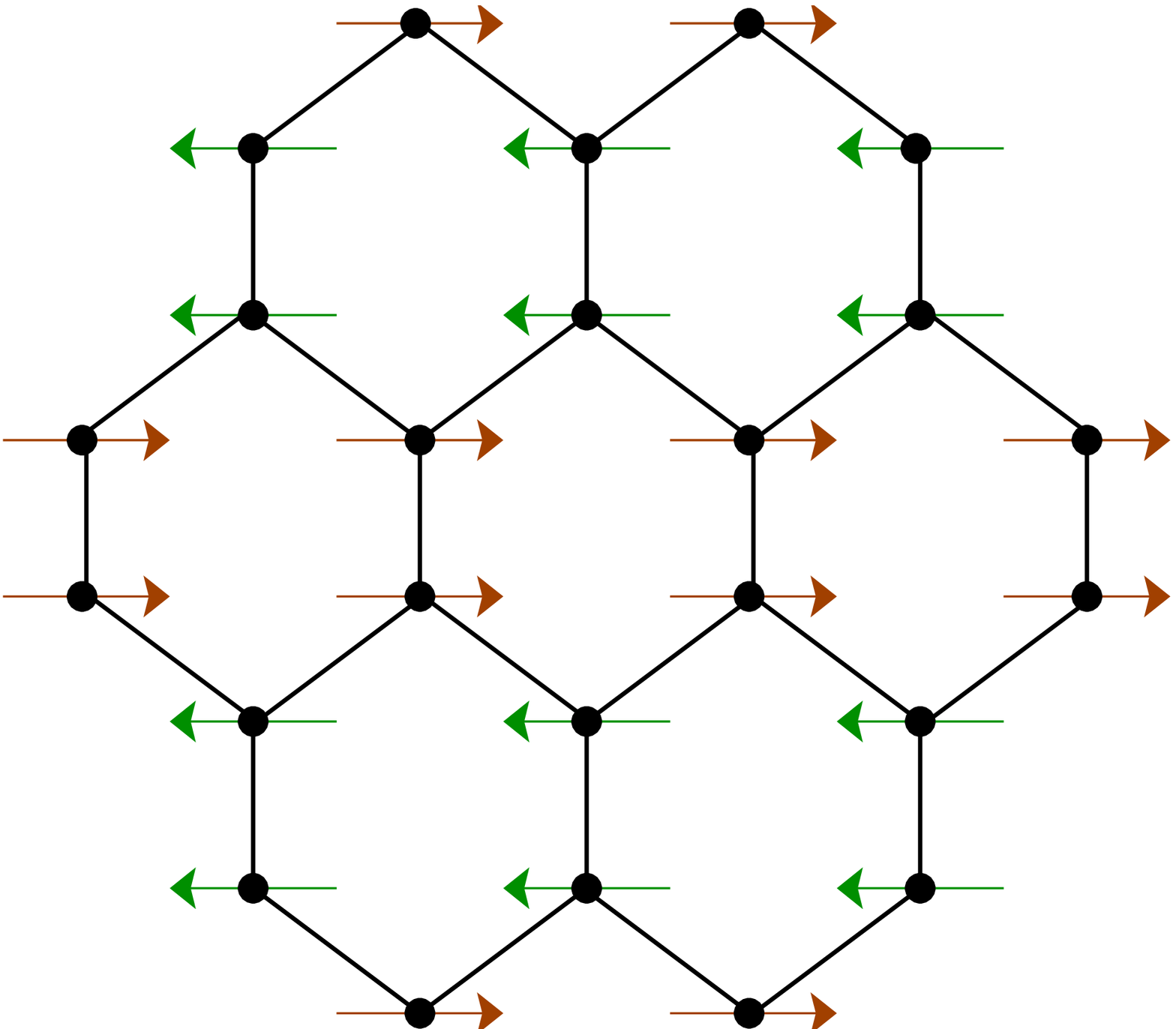}}}
}
\caption{(Color online) The $J_{1}$--$J_{2}$ $XY$ model on the honeycomb lattice with $J_{1}>0$ and $J_{2}>0$,
  showing (a) the bonds ($J_{1} \equiv$ ----- ; $J_{2} \equiv \textcolor{blue}{- - -}$), the two sites (\textcolor{red}{{\Large $\bullet$}}) A and B of the unit cell, and the Bravais lattice vectors $\hat{a}$ and $\hat{b}$; (b) the N\'{e}el planar, N(p), state; (c) the N\'{e}el $z$-aligned, N($z$), state; and (d) the N\'{e}el-II planar, N-II(p), state.  For the three states shown the arrows represent the directions of the spins located on lattice sites \textbullet.}
\label{model}
\end{center}
\end{figure*}

In view of these conflicting results for the spin-$\frac{1}{2}$
isotropic $XY$ model of Eq.\ (\ref{H_XY}) on the honeycomb lattice, it
seems worthwhile to apply another method to the model.  Since the CCM
has provided a consistent and accurate description for the closely
related spin-$\frac{1}{2}$ isotropic Heisenberg model of Eq.\
(\ref{H_H}), we now use it again here.  In Sec.\ \ref{model_sec} we
first describe the model itself in more detail, including its
classical counterpart, before reviewing the
basic ingredients of the CCM itself in Sec.\ \ref{ccm_sec} as it is
applied to general spin-lattice problems.  The results are then
presented in Sec.\ \ref{results_sec}, and we conclude with a summary
and discussion in Sec.\ \ref{summary_sec}.

\section{THE MODEL}
\label{model_sec}
In this paper we consider the Hamiltonian of Eq.\ (\ref{H_XY}) on the
honeycomb lattice for the case $s=\frac{1}{2}$.  We are interested in
the frustrated regime where $J_{1}>0$ and $J_{2} \equiv \kappa
J_{1}>0$.  The lattice and exchange bonds are illustrated in Fig.\
\ref{model}(a).  Henceforth we put $J_{1} \equiv 1$ to set the energy
scale.  The honeycomb lattice is a bipartite lattice, and the two
sites A and B of the unit cell are also shown in Fig.\ \ref{model}(a),
together with the two Bravais lattice vectors $\hat{a}=(1,0)$ and
$\hat{b}=(-\frac{1}{2},\frac{\sqrt3}{2})$, specified in terms of $x$
and $y$ coordinates in the lattice plane.  The honeycomb lattice
divides into two triangular sublattices $\mathcal{A}$ and
$\mathcal{B}$, such that A $\in \mathcal{A}$ and B $\in \mathcal{B}$.
We label the sites of the triangular Bravais lattice (i.e.,
$\mathcal{A}$) as ${\mathbf
  R}_{i}=m\hat{a}+n\hat{b}=(m-\frac{1}{2}n,\frac{\sqrt{3}}{2}n)$, and
consider the two sites (A and B) in the unit cell to have the same
value of ${\mathbf R}_{i}$.

The Wigner-Seitz unit cell is thus the parallelogram formed by the
lattice vectors $\hat{a}$ and $\hat{b}$, although it may equivalently,
and often more conveniently, be centered on a point of sixfold
rotational symmetry (i.e., at a center of any basic hexagon of the
lattice), in which case the unit cell is then bounded by the sides of
the hexagon.  The first Brillouin zone is then itself a hexagon that
is rotated by 90$^{\circ}$ with respect to the hexagonal Wigner-Seitz
cell.  A rotation in the plane by 180$^{\circ}$ around the center of
the hexagonal Wigner-Seitz cell (i.e., a 2D inversion) has the effect
of interchanging the two sublattices, ${\cal A} \rightarrow {\cal B}$.

We first summarize the results for the classical counterpart of the model, which pertain to both the
Heisenberg and $XY$ models of Eqs.\ (\ref{H_H}) and (\ref{H_XY})
respectively.  The most general state with coplanar order in the
$x_{s}y_{s}$ spin-coordinate plane can be described by a spiral wave
with wave vector ${\mathbf Q}$, together with an angle $\theta$ that
relates to the relative orientations of the two spins in the same unit
cell, both described by the same Bravais lattice vector ${\mathbf
  R}_{i}$.  Thus, we have
\begin{equation}
{\mathbf s}^{\mathcal A}_{i}=s[\cos({\mathbf Q}\cdot{\mathbf R}_{i})\hat{x}_{s}+\sin({\mathbf Q}\cdot{\mathbf R}_{i})\hat{y}_{s}]\,; \quad i \in {\mathcal A}\,, \label{s_A}
\end{equation}
for sites $i$ on the sublattice ${\mathcal A}$, and
\begin{equation}
{\mathbf s}^{\mathcal B}_{i}=-s[\cos({\mathbf Q}\cdot{\mathbf R}_{i}+\theta)\hat{x}_{s}+\sin({\mathbf Q}\cdot{\mathbf R}_{i}+\theta)\hat{y}_{s}]\,; \quad i \in {\mathcal B}\,, \label{s_B}
\end{equation}
for sites $i$ on the sublattice ${\mathcal B}$, where $\hat{x}_{s}$ and
$\hat{y}_{s}$ are orthogonal unit vectors in the $x_{s}y_{s}$ spin-coordinate plane, the normal direction to which is the $z_{s}$ axis, shown as the upward spin direction in Fig.\ \ref{model}(c).  The notation of Eqs.\
(\ref{s_A}) and (\ref{s_B}) is chosen such that the angle between NN
spins on the two sublattices with the same Bravais lattice vectors
${\mathbf R}_{i}$ is $\theta+\pi$, and hence such that the state with
N\'{e}el planar [N(p)] order shown in Fig.\ \ref{model}(b) is
described by ${\mathbf Q}={\mathbf \Gamma}\equiv(0,0)$ and $\theta=0$.

In this general coplanar spiral state described by Eqs.\ (\ref{s_A})
and (\ref{s_B}), the classical GS energy per spin is readily seen to
be given by
\begin{eqnarray}
\frac{E_{{\rm cl}}}{N}=&&-\frac{1}{2}J_{1}s^{2}[\cos\theta+\cos(\theta-Q_{b})+\cos(\theta-Q_{a}-Q_{b})] \nonumber \\
&&+J_{2}s^{2}[\cos Q_{a}+\cos Q_{b}+\cos(Q_{a}+Q_{b})]\,,  \label{E_cl}
\end{eqnarray}
where $Q_{a}\equiv{\mathbf Q}\cdot\hat{a}$ and $Q_{b}\equiv{\mathbf
  Q}\cdot\hat{b}$.  Clearly $Q_{a}$ and $Q_{b}$ are only uniquely defined
modulo $2\pi$ since, by definition, the addition of arbitrary
multiples of $2\pi$ to either (or both) quantity simply translates us
from the first Brillouin zone to another.  Minimization of Eq.\
(\ref{E_cl}) with respect to the three parameters $Q_{a},Q_{b}$ and
$\theta$ then yields the classical GS phase diagram.  One finds that
for small values of the frustration parameter in the range $0 < \kappa
< \frac{1}{6}$ the lowest-energy state is unique (up to rotations in the spin $x_{s}y_{s}$ plane), with collinear N(p)
AFM ordering and an energy per spin given by
\begin{equation}
\frac{E_{{\rm cl}}^{{\rm N(p)}}}{N}=\frac{3}{2}s^{2}(-1+2\kappa).  \label{E_np}
\end{equation}
with $J_{1}\equiv 1$.  Conversely, for $\kappa>\frac{1}{6}$, the
lowest-energy state is infinitely degenerate, with a wave vector that
is specified only by the single relation,
\begin{equation}
\cos Q_{a}+\cos Q_{b} + \cos(Q_{a}+Q_{b})=\frac{1}{2}\Big[\Big(\frac{1}{2\kappa}\Big)^{2}-3)\Big]\,.  \label{waveVec_gtrthan_1Over6}
\end{equation}
Thus, for $\kappa > \frac{1}{6}$, the
classically degenerate solutions form closed contours in the
reciprocal ${\mathbf Q}$-vector space.  For each point ($Q_{a},Q_{b}$)
on these contours the phase angle $\theta$ is uniquely specified by
the relations,
\begin{equation}
\sin \theta=2\kappa[\sin Q_{b}+\sin(Q_{a}+Q_{b})]\,,
\end{equation}
\begin{equation}
\cos \theta=2\kappa[1+\cos Q_{b}+\cos(Q_{a}+Q_{b})]\,.
\end{equation}
The GS energy per spin for these degenerate
spiral phases is given by
\begin{equation}
\frac{E^{{\rm spiral}}_{{\rm cl}}}{N}=-\frac{1}{2}s^{2}\Big(\frac{1}{4\kappa}+3\kappa\Big)\,,  \label{E_spiral_cl}
\end{equation}
with $J_{1}\equiv 1$.  By comparison of Eqs.\ (\ref{E_np}) and
(\ref{E_spiral_cl}) it is clear that there is a classical continuous
second-order transition at $\kappa_{{\rm cl}}=\frac{1}{6}$ between GS
phases with N(p) order for $\kappa < \kappa_{{\rm cl}}$ and spiral
planar [s(p)] order for $\kappa > \kappa_{{\rm cl}}$.

The classical spiral states themselves form two classes, one in the
range $\frac{1}{6}<\kappa<\frac{1}{2}$, and the other in the range
$\kappa>\frac{1}{2}$
\cite{Rastelli:1979_honey,Fouet:2001_honey,Mulder:2010_honey}.  For
the case $\frac{1}{6}<\kappa<\frac{1}{2}$ the manifold of classically
degenerate spiral wave vectors comprises closed contours around the
point ${\mathbf \Gamma}$.  At the critical point $\kappa=\frac{1}{2}$
this closed contour described by Eq.\ (\ref{waveVec_gtrthan_1Over6})
takes the form of a hexagon whose vertices are the six points
${\mathbf Q}={\mathbf M}^{(k)}$; $k=1,2,\cdots,6$, that are the
centers of the six edges of the hexagonal first Brillouin zone
corresponding to the honeycomb lattice.  Explicitly, these are given
by $(M_{a},M_{b})=(0,\pi)$, $(-\pi,\pi)$, $(-\pi,0)$, $(0,-\pi)$,
$(\pi,-\pi)$, and $(\pi,0)$, for the cases $k=1,2,\cdots,6$,
respectively.  Clearly there are only three distinct ${\mathbf M}$
vectors, which we denote as ${\mathbf M}^{\ast (l)}$; $l=1,2,3$.  They
can be chosen, for example, as $(M^{\ast (l)}_{a},M^{\ast
  (l)}_{b})=(0,\pi),(\pi,0),(\pi,\pi)$ for $l=1,2,3$, respectively,
which represent the midpoints of the three Brillouin zone boundaries
that join at a hexagonal vertex in ${\mathbf Q}$ space.  

For the case
$\kappa > \frac{1}{2}$, the classically degenerate spiral wave vectors
now comprise closed contours given by Eq.\
(\ref{waveVec_gtrthan_1Over6}), around the six points ${\mathbf
  Q}={\mathbf K}^{(k)}$; $k=1,2,\cdots,6$, that are the vertices of
the hexagonal first Brillouin zone.  These are given explicitly by
$(K_{a},K_{b})=(\frac{4\pi}{3},-\frac{2\pi}{3})$,
$(\frac{2\pi}{3},\frac{2\pi}{3})$, $(-\frac{2\pi}{3},\frac{4\pi}{3})$,
$(-\frac{4\pi}{3},\frac{2\pi}{3})$,
$(-\frac{2\pi}{3},-\frac{2\pi}{3})$, and
$(\frac{2\pi}{3},-\frac{4\pi}{3})$, respectively.  When $\kappa
\rightarrow \infty$, these closed contours collapse to the six points
${\mathbf Q}={\mathbf K}^{(k)}$, which are just the wave vectors
corresponding to the classical 120$^{\circ}$ ordering state on the
triangular lattice, just as expected in this limit where the two
sublattices of the honeycomb lattice become totally decoupled.
Clearly there are only two distinct ${\mathbf K}$ vectors, which we
denote as ${\mathbf K}^{\ast (n)}$; $n=1,2$.  They can be chosen as
$(K^{\ast (n)}_{a},K^{\ast
  (n)}_{b})=(-\frac{2\pi}{3},-\frac{2\pi}{3})$,
$(\frac{2\pi}{3},\frac{2\pi}{3})$ for $n=1,2$, respectively.  If we
define lattice vectors ${\mathbf d}_{1}=-\frac{1}{\sqrt{3}}\hat{y}$,
${\mathbf d}_{2}=\frac{1}{2}\hat{x}+\frac{1}{2\sqrt{3}}\hat{y}$, and
${\mathbf d}_{3}=-\frac{1}{2}\hat{x}+\frac{1}{2\sqrt{3}}\hat{y}$ to be
the vectors joining a B site on the ${\cal B}$ sublattice to its three
NN A sites on the ${\cal A}$ sublattice, then the six corners of the
first Brillouin zone at ${\mathbf Q}={\mathbf K^{(i)}}$ all have the
property that $({\mathbf K}^{(i)}\cdot {\mathbf d}_{1},\,{\mathbf
  K}^{(i)}\cdot {\mathbf d}_{2},\,{\mathbf K}^{(i)}\cdot {\mathbf
  d}_{3})$ is a permutation of
$(0,\,\frac{2\pi}{3},\,\frac{4\pi}{3})$.  The unit vectors joining the
six NN pairs of the same sublattice (i.e., the NNN pairs on the
honeycomb lattice) are $\pm \hat{b}_{l};\,l=1,2,3$, where
$\hat{b}_{1}\equiv {\mathbf d}_{2}-{\mathbf d}_{3}=\hat{a}$,
$\hat{b}_{2}\equiv {\mathbf d}_{3}-{\mathbf d}_{1}=\hat{b}$, and
$\hat{b}_{3}\equiv {\mathbf d}_{1}-{\mathbf d}_{2}=-\hat{a}-\hat{b}$.
The two distinct ${\mathbf K}^{\ast (n)}$ vectors then have the
property that ${\mathbf K}^{\ast (n)}\cdot
\hat{b}_{l}=(-1)^{n}\frac{2\pi}{3}$ for $l=1,2,3$.

As we mentioned in Sec.\ \ref{introd_sec}, it has been shown
\cite{Mulder:2010_honey} that in the region $\kappa > \frac{1}{6}$
$O(1/s)$ quantum corrections lift the huge classical GS degeneracy to
favor specific wave vectors, i.e., to select specific points from the
respective closed contours.  For the case $\frac{1}{6} < \kappa <
\frac{1}{2}$ the wave vectors so selected \cite{Mulder:2010_honey} are
those points on the corresponding closed contours around ${\mathbf
  \Gamma}$ that are intersected by the six vectors ${\mathbf
  Q}={\mathbf M}^{(k)}$.  For the limiting case $\kappa=\frac{1}{2}$
these are just the wave vectors ${\mathbf Q}={\mathbf M}^{(k)}$;
$k=1,2,\cdots,6$ themselves.  The three distinct such states, denoted
as ${\mathbf M}^{\ast (k)}$, $k=1,2,3$ above are precisely the three
distinct collinear N\'{e}el-II planar [N-II(p)] states, one of which,
namely that with ${\mathbf Q}={\mathbf M}^{\ast (1)}$, is illustrated in
Fig.\ \ref{model}(d).  They are all characterized by having
ferromagnetic NN bonds in one of the three lattice directions while those in
the remaining two directions are antiferromagnetic.  Finally, for the
case $\kappa>\frac{1}{2}$, the wave vectors selected by $O(1/s)$
quantum corrections \cite{Mulder:2010_honey} lie at the intersections
of the corresponding closed contours around the corner points
${\mathbf K}^{(k)}$ of the first Brillouin zone with the edges of the
zone.  As $\kappa$ increases from the value $\frac{1}{2}$ the selected
wave vectors thus move along the edges of the border zone, starting at
the center points ${\mathbf M}^{(k)}$ and moving monotonically
towards the vertices ${\mathbf K}^{(k)}$ as $\kappa \rightarrow
\infty$.

Our aim now is to study the $XY$ model of Eq.\ (\ref{H_XY}) on the honeycomb lattice for the case $s=\frac{1}{2}$.  We found previously \cite{Bishop:2012_honeyJ1-J2,RFB:2013_hcomb_SDVBC} that for the corresponding Heisenberg model of Eq.\ (\ref{H_H}) quantum
fluctuations are strong enough in the spin-$\frac{1}{2}$ case to
change the GS phase diagram very substantially from its classical
counterpart discussed above.  In particular, over the entire range $0
\leq \kappa \leq 1$, none of the spiral phases that are present
classically for all values $\kappa > \frac{1}{6}$ form the stable GS
phase for the spin-$\frac{1}{2}$ system.  This finding is a
particularly dramatic confirmation of the more general observation
that quantum fluctuations tend to favor GS phases with collinear
ordering over those, such as the spiral phases, with noncollinear
order.  Similarly, we also found for the spin-$\frac{1}{2}$ Heisenberg
model that the N\'{e}el state is stabilized out to a value
$\kappa_{c_{1}} \approx 0.21$, substantially beyond the corresponding
classical value $\kappa_{{\rm cl}}=\frac{1}{6}$.  Furthermore, we
found that the collinear N\'{e}el-II state, which exists as the stable
GS phase for the classical $J_{1}$--$J_{2}$ Heisenberg model on the
honeycomb lattice only at the highly degenerate point
$\kappa=\frac{1}{2}$, provides the stable GS phase for the
corresponding spin-$\frac{1}{2}$ model for values
$\kappa>\kappa_{c_{3}}\approx 0.65 \pm 0.05$ within the window $0 \leq
\kappa \leq 1$ that was examined.  Over the entire range $\kappa >
\kappa_{c_{2}} \approx 0.385 \pm 0.005$, we found a very close
competition between the N\'{e}el-II state and the SDVBC (or lattice
nematic) state.  These two states are closely related, and indeed the
latter is obtained from the former by replacing all of the NN
ferromagnetic spin pairs by spin-zero dimers.  Hence the two states
break the lattice rotational symmetry in the same way.  Results of our
CCM analysis \cite{RFB:2013_hcomb_SDVBC} provided good numerical
evidence that the SDVBC state is favored as the GS phase in the region
$\kappa_{c_{2}} < \kappa < \kappa_{c_{3}}$.  Finally, in the region
$\kappa_{c_{1}} < \kappa < \kappa_{c_{2}}$, our previous CCM analysis
\cite{Bishop:2012_honeyJ1-J2,RFB:2013_hcomb_SDVBC} strongly indicated
that the stable GS phase is the PVBC state, which preserves the
lattice rotational symmetry.

In view of the above results, obtained from analyses using the CCM, we
now also apply the CCM to the corresponding spin-$\frac{1}{2}$
$J_{1}$--$J_{1}$ $XY$ model of Eq.\ (\ref{H_XY}) on the honeycomb
lattice.  Since the two models of Eqs.\ (\ref{H_H}) and (\ref{H_XY})
share exactly the same classical GS phase diagram (i.e., in the limit
as $s \rightarrow \infty$) on the honeycomb lattice, it is of
particular interest to enquire whether quantum fluctuations behave
differently for the two models in the case $s=\frac{1}{2}$, where
their effects are expected to be largest.

As a final point we recall that quantum spin-$\frac{1}{2}$ operators can be mapped exactly \cite{Matsubara:1956_spinHalf-operator} onto hard-core (HC) boson operators by making the identifications,
\begin{equation}
s^{+}_{k} \rightarrow b^{\dagger}_{k}\,; \quad s^{-}_{k} \rightarrow b_{k}\,; \quad s^{z}_{k} \rightarrow n_{k}-\frac{1}{2}\,, \label{HC_bosonOper}
\end{equation}
where $s^{\pm}_{k}\equiv s^{x}_{k} \pm is^{y}_{k}$ are the usual spin
raising and lowering operators respectively, $b^{\dagger}_{k}$ and $b_{k}$
are HC boson creation and annihilation operators at site $k$, and
$n_{k} \equiv b^{\dagger}_{k}b_{k}$ is the boson number operator at
site $k$.  The imposition of the HC constraint that no more than one
boson can occupy any site (viz., $n_{k}=0,1$ only) then guarantees
that the bosonic Hilbert space has the same dimensionality as that of
the spin-$\frac{1}{2}$ system (viz., two states per site).  For the
bosons we assume that operators at different sites ($k \neq l$)
commute as usual,
\begin{equation}
[b^{\dagger}_{k},b^{\dagger}_{l}]=[b_{k},b_{l}]=[b_{k},b^{\dagger}_{l}]=0\,; \quad k \neq l\;, \label{commRel_diffSites}
\end{equation}
but to exclude multiple occupancy at any site we assume fermion-like anticommutation relations when $k=l$,
\begin{equation}
\{b_{k},b_{k}\}=0=\{b^{\dagger}_{k},b^{\dagger}_{k}\}\,; \quad \{b^{\dagger}_{k},b_{k}\}=1\,.  \label{fermion_anticomm}
\end{equation}

It is easy to show from Eq.\ (\ref{fermion_anticomm}) that these HC bosons obey the same commutation relations with the number operator,
\begin{equation}
[n_{k},b^{\dagger}_{k}]=b^{\dagger}_{k}\,; \quad [n_{k},b_{k}]=-b_{k}\,, \label{commRel_numOper}
\end{equation}
as do ordinary bosons [i.e., those obeying the usual commutation
relations rather than their anticommutation counterparts in Eq.\
(\ref{fermion_anticomm})].  Furthermore, Eq.\ (\ref{fermion_anticomm})
also readily implies the commutation relation for HC bosons,
\begin{equation}
[b^{\dagger}_{k},b_{k}]=2n_{k}-1\,.  \label{cummRelatn_HCBosons}
\end{equation}
Thus, one observes from Eqs.\ (\ref{commRel_diffSites}),
(\ref{commRel_numOper}) and (\ref{cummRelatn_HCBosons}) that the
mapping of Eq.\ (\ref{HC_bosonOper}) produces the correct SU(2) spin
commutation relations,
\begin{equation}
[s^{\mu}_{k},s^{\nu}_{l}]=0 \quad \forall\, \mu,\nu \, \in \{+,-,z\}\,; \quad k \neq l\,, \label{SU2_commRel_diffSites}
\end{equation}
for spins on different sites, and
\begin{equation}
[s^{z}_{k},s^{\pm}_{k}]=\pm s^{\pm}_{k}\,; \quad [s^{+}_{k},s^{-}_{k}]=2s^{z}_{k}\,, \label{commRel_sameSites}
\end{equation}
for operators on the same site.  The commutation relations of Eq.\
(\ref{commRel_sameSites}) readily yield the anticommutation relation,
\begin{equation}
\{s^{+}_{k},s^{-}_{k}\}=2s(s+1)-2s^{2}_{z}\,,
\end{equation}
which, for the special case $s=\frac{1}{2}$ only, gives
\begin{equation}
\{s^{+}_{k},s^{-}_{k}\}=1\,; \quad s=\frac{1}{2}\,.  \label{anticommRel_spinHalf}
\end{equation}

Equation (\ref{anticommRel_spinHalf}), together with the corresponding
results $(s^{+})^{2}=0=(s^{-})^{2}$ that hold for $s=\frac{1}{2}$,
then exactly reproduce Eq.\ (\ref{fermion_anticomm}) under the mapping
of Eq.\ (\ref{HC_bosonOper}).

This exact isomorphism between HC bosonic systems and quantum
spin-$\frac{1}{2}$ systems on a lattice can often be used to gain
additional insight into either system.  The spin-$\frac{1}{2}$
$J_{1}$--$J_{2}$ $XY$ model of Eq.\ (\ref{H_XY}) maps onto the
particularly revealing HC boson model,
\begin{equation}
{\cal H}_{{XY}}=\frac{J_{1}}{2}\sum_{\langle i,j \rangle}(b^{\dagger}_{i}b_{j}+b^{\dagger}_{j}b_{i})+\frac{J_{2}}{2}\sum_{\langle \langle i,k \rangle \rangle}(b^{\dagger}_{i}b_{k}+b^{\dagger}_{k}b_{i})\,,  \label{H_XY_BH}
\end{equation} 
which is simply a model of frustrated hopping, involving a competition
between hopping between NN and NNN pairs of sites.

In this context it is interesting to compare the above model with
another archetypal model of lattice bosons, namely the Bose-Hubbard
(BH) model, the Hamiltonian for which may be written in the usual
form,
\begin{equation}
{\cal H}_{{\rm BH}}=-t_{1}\sum_{\langle i,j \rangle}(b^{\dagger}_{i}b_{j}+b^{\dagger}_{j}b_{i})+\frac{U}{2}\sum_{i}n_{i}(n_{i}-1)-\mu \sum_{i}n_{i}\,,  \label{H_BH}
\end{equation} 
where $t_{i}$ is the NN hopping parameter, $U$ is the on-site
(Coulomb) interaction parameter (repulsive if $U>0$), and $\mu$ is the
chemical potential.  In Eq.\ (\ref{H_BH}) the bosons obey the usual
commutation rules of Eq.\ (\ref{commRel_diffSites}) but with those of
Eq.\ (\ref{fermion_anticomm}) where the anticommutators are replaced
by commutators.  Clearly, in the HC limit $U \rightarrow \infty$, the
second term in Eq.\ (\ref{H_BH}) forces each site to contain no more
than one boson, and this term may be removed, but with the HC
constraint ($n_{i}=0,1$ only) imposed in its place.  As we have seen,
this constraint may be imposed by the HC boson operator relations of
Eqs.\ (\ref{commRel_diffSites}) and (\ref{fermion_anticomm}).  

We also note that the sign of the NN hopping parameter $t_{1}$ in Eq.\ (\ref{H_BH}) is irrelevant
since the honeycomb lattice is bipartite and the operator algebra
relations are invariant under the replacement $b_{i} \rightarrow
-b_{i}$, as we now show explicitly.  Thus, by making use of Eq.\ (\ref{commRel_numOper}) and the nested commutator expansion for operator products of the form $e^{-A}Be^{A}$, it is easy to show that the unitary operator,
\begin{equation}
\hat{U}\equiv \exp\Big(i \pi \sum_{k \in {\cal A}} n_{k}\Big)\,,
\end{equation}
has the following mode of action on the basic HC boson operators,
\begin{equation}
\hat{U}^{\dagger}b_{l}\hat{U}=\epsilon_{l}b_{l}\,; \quad \hat{U}^{\dagger}b_{l}^{\dagger}\hat{U}=\epsilon_{l}b_{l}^{\dagger}\,,
\end{equation}
where
\begin{equation}
\epsilon_{l} = \left\{ 
\begin{array}{l l}
  -1\,; & \quad l \in {\cal A}\,,\\
  +1\,;  & \quad l \in {\cal B}\,.\\ 
\end{array} \right. 
\end{equation}
Under this unitary transformation, ${\cal H}_{{\rm BH}}$ is mapped as follows,
\begin{equation}
\hat{U}^{\dagger}{\cal H}_{{\rm BH}}(t_{1},U,\mu)\hat{U} = {\cal H}_{{\rm BH}}(-t_{1},U,\mu)\,.
\end{equation}
Since $\hat{U}$ is unitary both ${\cal H}_{{\rm BH}}(t_{1},U,\mu)$ and ${\cal H}_{{\rm BH}}(-t_{1},U,\mu)$ have the same energy spectrum and there is a one-to-one correspondence between their eigenstates.  Similarly, under $\hat{U}$, ${\cal H}_{{XY}}$ of Eq.\ (\ref{H_XY_BH}) gets mapped as follows,
\begin{equation}
\hat{U}^{\dagger}\hat{{\cal H}}_{{XY}}(J_{1},J_{2})\hat{U}={{\cal H}}_{{XY}}(-J_{1},J_{2})\,,
\end{equation}
which shows that the sign of $J_{1}$ is irrelevant for this Hamiltonian.

The analog of this for the isomorphic spin-lattice model is
to make a rotation of 180$^{\circ}$ about the $z_{s}$ axis of the spins on
one of the two sublattices, which cannot change the physics since it
simply corresponds to a different choice of direction of the local
axes of spin quantization.  This has the effect, however, of replacing
$s^{x}_{i} \rightarrow -s^{x}_{i}$, $s^{y}_{i} \rightarrow -s^{y}_{i}$
on one sublattice, which in turn shows that the replacement $J_{1}
\rightarrow -J_{1}$ does not change the physics.  As indicated
previously we choose $J_{1} \equiv +1$ to set the energy scale (and
consider the case of frustration where $J_{2}>0$).

From the mapping of Eq.\ (\ref{HC_bosonOper}) one sees that the role
of the chemical potential $\mu$ in the BH model for lattice bosons is
played by an applied magnetic field in the $z_{s}$ direction ($\mu
\rightarrow B_{z}$) for the equivalent lattice spin model.  In our case,
where we consider $B_{z}=0$, we see that the BH Hamiltonian of Eq.\
(\ref{H_BH}) thus reduces in the HC limit to precisely the first term
of ${\cal H}_{{XY}}$ where $t_{1} \rightarrow \frac{1}{2}J_{1}$.

We note that ${\cal H}_{{XY}}$ of Eq.\ (\ref{H_XY}) commutes with the
total lattice magnetization in the $z_{s}$ direction, ${\cal M} \equiv
\sum^{N}_{k=1}s^{z}_{k}$, where $N$ is the number of lattice sites,
and $s^{z}_{k}$ now refers to a global set of spin axes rather than
the local axes used later to define the order parameter $M$ in Sec.\
\ref{ccm_sec}.  For the isomorphic boson model of Eq.\ (\ref{H_XY_BH})
the analogous statement is that $H_{{XY}}$ commutes with the total
boson number.  For bipartite lattices with AFM interactions the GS
phases are expected to lie in the ${\cal M}=0$ sector, and the mapping
of Eq.\ (\ref{HC_bosonOper}) thus shows that for the analogous boson
lattice model we are interested in the corresponding case of
half-filling, $n \equiv \langle n_{k} \rangle = \frac{1}{2}$.

In this context it is interesting to note that one of the
characteristic features of HC bosons on a lattice with NN hopping
augmented by additional off-site two-body interactions is the
appearance of solid phases at half-filling ($n=\frac{1}{2}$) with
either a charge-density wave (CDW) or a bond-order wave (BOW) ordering
\cite{Voit:1992_BOW,MNakamura:1999_BOW,Nakamura:2000_BOW}.  Whereas a
CDW at half-filling is characterized by long-range order (LRO) in
fluctuations in the density operator at site $k$,
\begin{equation}
{\cal O}^{{\rm CDW}}_{k} \equiv a\epsilon_{k}\Big(b^{\dagger}_{k}b_{k}-\frac{1}{2}\Big)\,,
\end{equation}
where $\epsilon_{k}$ takes opposite value on the two sublattices
(viz., $\epsilon_{k} \equiv -1$ if $k \in {\cal A}$ and
$\epsilon_{k} \equiv +1$ if $k \in {\cal B}$), a BOW is
characterized by LRO in fluctuations in the hopping (or kinetic
energy) operator at site $k$,
\begin{equation}
{\cal O}^{{\rm BOW}}_{k}=(1+a \epsilon_{k}) \sum_{k'}(b^{\dagger}_{k}b_{k'}+b^{\dagger}_{k'}b_{k})\,,
\end{equation}
where the sum over $k'$ runs over the NN sites to $k$.  Both the CDW
and BOW states thus break lattice translational symmetry.  In addition
the CDW state breaks particle-hole symmetry, whereas the BOW state
breaks inversion symmetry.

A typical NN two-body interaction term, $V \sum_{\langle i,j
  \rangle}n_{i}n_{j}$ in the bosonic system, maps into a corresponding
term $V \sum_{\langle i,j \rangle}s^{z}_{i}s^{z}_{j}$ in the
isomorphic spin-lattice problem, and thus replaces the NN isotropic $XY$
interaction in the first term of $H_{{XY}}$ of Eq.\ (\ref{H_XY}) by an
effective {\it XXZ} interaction.  By contrast, in the present paper the NN $XY$
interaction is frustrated by a NNN interaction of the same form.
Thus, in the bosonic language, the frustration is caused not by
off-site two-body interactions but by the introduction of NNN hopping
between sites of the same sublattice competing with NN hopping between
sites of different sublattices.  It is then interesting to speculate
whether such solid phases as those with CDW ordering might also occur
in this case.

Clearly a CDW phase in the boson lattice system corresponds to a
N\'{e}el $z$-aligned phase of the sort illustrated in Fig.\
\ref{model}(c), due to the exact mapping of Eq.\ (\ref{HC_bosonOper}).
Interestingly, this is precisely the GS stable phase identified for the
spin-$\frac{1}{2}$ $J_{1}$--$J_{2}$ $XY$ model on the honeycomb lattice
in the recent paper \cite{Zhu:2013_honey_XY} discussed in Sec.\
\ref{introd_sec}.  For this reason we specifically include it as a
candidate model state in our CCM study of this model, as discussed in
Sec.\ \ref{ccm_sec}.

Finally, we note that the Hamiltonian of Eq.\ (\ref{H_XY_BH}) on the
honeycomb lattice is very similar to one discussed by Haldane
\cite{Haldane:1988_honeycomb} for noninteracting electrons on a honeycomb lattice
in which the electronic spin degrees of freedom are suppressed, but where both NN
and NNN hopping terms are included.  The NN hopping parameter $t_{1}$
is considered real, as is the case here, but the NNN hopping parameter
$t_{2}$ becomes complex, $t_{2} \rightarrow |t_{2}|{\rm e}^{i \phi}$
in one direction (and hence $t_{2} \rightarrow |t_{2}|{\rm e}^{-i
  \phi}$ in the other).  Haldane shows that such a phase factor $\phi$
can be induced for $t_{2}$, leaving $t_{1}$ real, by the imposition of
a suitable magnetic field in the $z$ direction perpendicular to the 2D
honeycomb plane.  He then shows that when $\phi \neq 0,\pi$ the model
realizes a topological insulator phase (viz., an integer quantum Hall
state) in its $T=0$ GS phase diagram, apart from the Mott insulator
phase that holds for $\phi = 0,\pi$.  Clearly, the Haldane model
\cite{Haldane:1988_honeycomb} is simply related to our ${\cal H}_{{XY}}$ model if
the spinless fermions are replaced by HC bosons and $\phi \rightarrow
\pi$.  This provides yet another justification for expecting that the
$T=0$ GS phase diagram of ${\cal H}_{{XY}}$ on the honeycomb lattice might
hold surprises in store.

In Sec.\ \ref{ccm_sec} we now describe the CCM technique employed
here, together with the choice of model states suggested by our
discussion of the model above.

\section{THE COUPLED CLUSTER METHOD}
\label{ccm_sec}
The CCM has been very successfully employed for a wide variety of
quantum many-body systems (see, e.g., Refs.\
\onlinecite{Bishop:1987_ccm,Arponen:1991_ccm,Bishop:1991_TheorChimActa_QMBT,Bishop:1998_QMBT_coll,Fa:2004_QM-coll}),
where it typically yields results of accuracy better than or
comparable to those of alternative techniques.  In recent years it has
been widely applied, for example, to investigate the GS phase
structure of many spin-lattice models, especially 2D ones, of interest
in quantum magnetism (see, e.g., Refs.\
\onlinecite{DJJF:2011_honeycomb,PHYLi:2012_honeycomb_J1neg,Bishop:2012_honey_phase,Bishop:2012_honeyJ1-J2,RFB:2013_hcomb_SDVBC,Bishop:1991_XXZ_PRB44,Zeng:1998_SqLatt_TrianLatt,Kruger:2000_JJprime,Bishop:2000_XXZ,Fa:2001_trian-kagome,Darradi:2005_Shastry-Sutherland,Schm:2006_stackSqLatt,Bi:2008_PRB_J1xxzJ2xxz,Bi:2008_JPCM_J1J1primeJ2,Darradi:2008_J1J2mod,Bi:2009_SqTriangle,Richter2010:J1J2mod_FM,Bishop:2010_UJack,Bishop:2010_KagomeSq,Reuther:2011_J1J2J3mod,Gotze:2011_kagome,Bishop:2012_checkerboard,Li:2012_honey_full,Li:2012_anisotropic_kagomeSq,Li:2013_chevron,Bishop:2013_crossStripe,Fa:2004_QM-coll,Darradi:2009_J1J2_XXZmod},
and references cited therein).  The general consensus is that it now
provides one of the most accurate methods available to study such
models.  In particular, the CCM offers a systematic technique for the
investigation of various possible GS phases, including an accurate
determination of the associated QCPs that demarcate their regions of
stability.  Very importantly, since it is formulated with well-defined
hierarchies of approximations, which we describe below, it is capable
of systematic improvement in accuracy, albeit at an increasing
computational cost.

The CCM, as we explain below, offers a size-extensive method in which
we work from the outset in the thermodynamic (i.e., infinite-lattice) limit ($N \rightarrow
\infty$).  Hence, no finite-size scaling is ever required.  However,
what it does require is for us to select a suitable, normalized model (or
reference) state $|\Phi\rangle$, with respect to which the quantum
correlations in the exact GS phase under study may then be explicitly
included, in principle exactly, via a correlation operator, $S$,
chosen as described below (and see e.g., Refs.\
\onlinecite{Zeng:1998_SqLatt_TrianLatt,Bishop:1991_TheorChimActa_QMBT,Bishop:1998_QMBT_coll,Fa:2004_QM-coll},
and references cited therein).  In the present study we use,
separately, each of the N(p), N($z$), and N-II(p) states shown in
Figs.\ \ref{model}(b)-(d), respectively, as our choices for CCM model
states.  Each of these states is characterized as being an
independent-spin product state, in which the choice of state for the
spin on every site is formally independent of the choice for that of
all others.

In order to make the subsequent computational implementation of the
technique as universal (i.e., as independent of the choice of model
state) as possible, so that every lattice site can then be treated on
an equal basis, we then choose local coordinate frames in spin space
for each model state we employ.  The choice is made so that on each
lattice site in each model state the spin aligns along, say, the
negative $z_{s}$ axis (henceforth called the downward) direction in its own
spin-coordinate frame.  Such canonical passive rotations leave the
SU(2) commutation relations unchanged, and hence have no physical
consequences.

We denote the exact, fully correlated, GS ket- and bra-state wave
functions of the phase of the system under study as $|\Psi\rangle$ and
$\langle\tilde{\Psi}|$, respectively.  They satisfy the respective GS
Schr\"{o}dinger equations,
\begin{equation}
H|\Psi\rangle = E|\Psi\rangle\,; \quad \langle\tilde{\Psi}|H=E\langle\tilde{\Psi}|\,,
\end{equation}
and their normalizations are chosen so that
$\langle\tilde{\Psi}|\Psi\rangle = \langle{\Phi}|\Psi\rangle =
\langle{\Phi}|\Phi\rangle = 1$.  The CCM then employs the distinctive
exponential parametrizations of the exact ket and bra states with
respect to the corresponding chosen model state, as
\begin{equation}
|\Psi\rangle=e^{S}|\Phi\rangle\,; \qquad \langle\tilde{\Psi}|=\langle\Phi|\tilde{S}e^{-S}\,, \label{ket_bra_eq}
\end{equation}  
which are one of the hallmarks of the method.  The two correlation operators, $S$ and $\tilde{S}$ are now decomposed formally as
\begin{equation}
  S=\sum_{I \neq 0}{\cal S}_{I}C^{+}_{I}\,; \qquad \tilde{S}=1+\sum_{I \neq 0}\tilde{{\cal S}}_{I}C^{-}_{I}\,,  \label{S_operators}
\end{equation}
where we define $C^{+}_{0} \equiv 1$ to be the identity operator, and
where the set-index $I$ represents a particular set of lattice sites.
It is used, as we discuss below, to denote a multispin-flip
configuration with respect to the model state $|\Phi\rangle$, such
that $C^{+}_{I}|\Phi\rangle$ represents the corresponding wave
function for this configuration of spins.  The operators $C^{+}_{I}$
and $C^{-}_{I} \equiv (C^{+}_{I})^{\dagger}$ are thus creation
and destruction operators, respectively, with respect to the
model state $|\Phi\rangle$ considered as a generalized vacuum state.
They are chosen so as to obey the corresponding relations,
\begin{equation}
\langle\Phi|C^{+}_{I}=0=C^{-}_{I}|\Phi\rangle\,; \quad \forall I \neq 0\,.
\end{equation}
The set $\{C^{+}_{I}\}$ thus forms a mutually commuting, complete set of
multispin creation operators with respect to the model state
$|\Phi\rangle$ as the corresponding cyclic vector.

We now discuss the choice of set-indices $\{I\}$ and creation
operators $\{C^{+}_{I}\}$ in more detail for the specific case of
spin-lattice models considered here.  With our choice of local
spin-coordinate frames as descried above, in which any
independent-spin product model state has the universal form
$|\Phi\rangle=|\downarrow\downarrow\downarrow\cdots\downarrow\rangle$
with all spins pointing down, the operators $C^{+}_{I}$ also take a
universal form.  Thus, $C^{+}_{I} \rightarrow
s^{+}_{l_{1}}s^{+}_{l_{2}}\cdots s^{+}_{l_{n}}$, a product of
single-spin raising operators, $s^{+}_{l} \equiv
s^{x}_{l}+is^{y}_{l}$, where the set-index $I \rightarrow
\{l_{1},l_{2},\cdots , l_{n};\; n=1,2,\cdots , 2sN\}$, a set of
(possibly repeated) lattice site indices, where $N (\rightarrow
\infty)$ is the total number of sites.  Clearly, in the general case,
for arbitrary spin quantum number $s$, a spin raising operator
$s^{+}_{l}$ can be applied no more than $2s$ times on a given site
$l$.  Therefore, a given site-index $l$ may appear no more than 2$s$
times in any set-index $I$ included in the sums in Eq.\
(\ref{S_operators}).  Thus, for our present study, where
$s=\frac{1}{2}$, each site-index $l_{k}$ included in any 
set-index $I$ may appear no more than once.

The CCM correlation coefficients, $\{{\cal S}_{I},{\tilde{\cal
    S}}_{I}\}$, in terms of which an arbitrary GS property may
formally be expressed, are themselves now calculated by minimizing the
GS energy expectation functional,
\begin{equation}
\bar{H}=\bar{H}\{{\cal S}_{I},{\tilde{\cal S}_{I}}\}\equiv \langle\Phi|\tilde{S}e^{-S}He^{S}|\Phi\rangle\,,  \label{GS_E_exp_funct}
\end{equation}
with respect to each of the coefficients $\{\tilde{\cal S}_{I},{\cal S}_{I};\,\forall I \neq 0\}$.  Equations (\ref{S_operators}) and (\ref{GS_E_exp_funct}) thus yields 
the coupled sets of equations,
\begin{equation}
\langle\Phi|C^{-}_{I}e^{-S}He^{S}|\Phi\rangle=0\,; \quad \forall I \neq 0\,,  \label{ccm_ket-eq}
\end{equation}
by minimization with respect to $\tilde{\cal S}_{I}$, and
\begin{equation}
\langle\Phi|\tilde{S}e^{-S}[H,C^{+}_{I}]e^{S}|\Phi\rangle = 0\,; \quad \forall I \neq 0\,, \label{ccm_bra-eq}
\end{equation}
by minimization with respect to ${\cal S}_{I}$.  Once Eq.\
(\ref{ccm_ket-eq}) is satisfied, the GS energy, which is the value of
$\bar{H}$ at the minimum, may be simply expressed as
\begin{equation}
E=\langle\Phi|e^{-S}He^{S}|\Phi\rangle\,,  \label{gs_E}
\end{equation}
and Eq.\ (\ref{ccm_bra-eq}) may be written equivalently as
\begin{equation}
\langle\Phi|\tilde{S}(e^{-S}He^{S}-E)C^{+}_{I}|\Phi\rangle = 0\,; \quad \forall I \neq 0\,. \label{gs_E2}
\end{equation}

The CCM equations (\ref{ccm_ket-eq}) for the ket-state correlation
coefficients $\{{\cal S}_{I}\}$ comprise coupled sets of nonlinear
equations, due to the presence of the operator $S$ in the exponentials.  However,
a key feature of the CCM is that in the equations we utilize for
solution, $S$ only appears in the combination $e^{-S}He^{S}$, a
similarity transform of the Hamiltonian.  In turn, this form may be
expanded in terms of the well-known nested commutator sum.  Another
key feature of the CCM is that this otherwise infinite sum actually
terminates {\it exactly} with the double commutator term, due firstly to the
fact that all of the terms in Eq.\ (\ref{S_operators}) comprising $S$
mutually commute and are simple products of spin-raising operators, as
described above, and secondly to the basic SU(2) commutator relations
(and see, e.g., Refs.\
\onlinecite{Zeng:1998_SqLatt_TrianLatt,Fa:2001_trian-kagome} for further
details).  Similar exact terminations occur for the GS expectation
value of any physical operator, such as the magnetic order parameter,
which is defined to be the average local on-site magnetization,
\begin{align}
M & \equiv -\frac{1}{N}\langle\tilde{\Psi}|\sum_{k=1}^{N}s^{z}_{k}|\Psi\rangle \nonumber \\
  & = -\frac{1}{N}\langle\tilde{\Phi}|\tilde{S}\sum_{k=1}^{N}e^{-S}s^{z}_{k}e^{S}|\Phi\rangle \,,  \label{gs_M}
\end{align}
where $s^{z}_{k}$ is defined with respect to the local rotated
spin-coordinate frame on each lattice site $k$, as described above, and as opposed to the global spin-coordinate frame used to define the total lattice magnetization ${\cal M}$ in Sec.\ \ref{model_sec}.

We note too that the CCM parametrizations of Eqs.\ (\ref{ket_bra_eq}) and (\ref{S_operators}),
which imply, for example, that every term in $S$ commutes with all of
the others, together with the nested commutator expansion for
$e^{-S}He^{S}$ in Eq.\ (\ref{ccm_ket-eq}), suffice completely to show
rather readily \cite{Bishop:1998_QMBT_coll} that the CCM exactly obeys
the Goldstone linked cluster theorem at {\it any} level of approximation
involving truncations on the index-set $\{I\}$ retained in the sums in
Eq.\ (\ref{S_operators}).  Similarly, one can also show
\cite{Bishop:1998_QMBT_coll} that the CCM obeys the important
Hellmann-Feynman theorem at all such levels of approximation.

Once an approximation has been made as to which multispin-flip
configurations $\{I\}$ to retain in the expansions of Eq.\
(\ref{S_operators}) for the CCM correlation operators $S$ and
$\tilde{S}$, no further approximation is made.  The set of nonlinear
equations (\ref{ccm_ket-eq}) for the coefficients $\{{\cal S}_{I}\}$
is first solved.  They are then used as input to solve the linear set
of equations (\ref{ccm_bra-eq}) or (\ref{gs_E2}) for the coefficients
$\{\tilde{{\cal S}}_{I}\}$.  Any GS expectation value may then be
exactly computed at the same level of approximation.  In this work we
use the well-tested localized (lattice-animal-based subsystem) LSUB$m$
truncation scheme
\cite{Zeng:1998_SqLatt_TrianLatt,Kruger:2000_JJprime,Bishop:2000_XXZ,Fa:2001_trian-kagome,Darradi:2005_Shastry-Sutherland,Schm:2006_stackSqLatt,Bi:2008_PRB_J1xxzJ2xxz,Bi:2008_JPCM_J1J1primeJ2,Darradi:2008_J1J2mod,Bi:2009_SqTriangle,Richter2010:J1J2mod_FM,Bishop:2010_UJack,Bishop:2010_KagomeSq,Reuther:2011_J1J2J3mod,Gotze:2011_kagome,Bishop:2012_checkerboard,Li:2012_honey_full,Li:2012_anisotropic_kagomeSq,Li:2013_chevron,Bishop:2013_crossStripe,Fa:2004_QM-coll,Darradi:2009_J1J2_XXZmod},
which has been applied with considerable success to many different 2D
spin-lattice models.  At the $m$th level of approximation in the
LSUB$m$ scheme we retain all multispin-flip configurations $\{I\}$
defined over $m$ or fewer contiguous lattice sites.  The
configurations of spins (or clusters) retained are defined to be
contiguous when every site in the configuration is adjacent (i.e., as
a NN) to at least one other site in the configuration.  The interested
reader is referred to the literature (see, e.g., Ref.\
\onlinecite{Zeng:1998_SqLatt_TrianLatt}) for specific examples that
illustrate the LSUB$m$ scheme in some detail.

Even after all space- and point-group symmetries of the lattice and
the CCM model state used are taken fully into account, the number
$N_{f}$ of such distinct fundamental configurations that are retained
in the LSUB$m$ scheme grows rapidly with increasing values of the
truncation index $m$.  It thus becomes necessary to use massive
parallelization together with supercomputing resources
\cite{Zeng:1998_SqLatt_TrianLatt,ccm} for the higher-order
approximations.  In the present case we have been able to perform CCM
calculations up to the LSUB10 level based on the N($z$) model state,
and up to the LSUB12 level based on both the N(p) and N-II(p) model
states.  For example, for the N-II(p) state the number of fundamental
configurations at the LSUB12 level is $N_{f}=818300$.

Since we work from the outset in the infinite-lattice ($N \rightarrow
\infty$) limit, the only extrapolation we need finally to make is to take
the $m \rightarrow \infty$ limit in the LSUB$m$ truncation index, where our
results are, in principle, {\it exact}, since we make no other
approximations.  For example, the LSUB$m$ values for the GS energy per
spin, $E(m)/N$, always converge very rapidly, and we use the very
well-tested extrapolation scheme
\cite{DJJF:2011_honeycomb,Kruger:2000_JJprime,Bishop:2000_XXZ,Fa:2001_trian-kagome,Darradi:2005_Shastry-Sutherland,Schm:2006_stackSqLatt,Bi:2008_JPCM_J1J1primeJ2,Bi:2008_PRB_J1xxzJ2xxz,Darradi:2008_J1J2mod,Richter2010:J1J2mod_FM,Reuther:2011_J1J2J3mod,Bishop:2012_checkerboard,Li:2012_anisotropic_kagomeSq,RFB:2013_hcomb_SDVBC,Li:2013_chevron,Bishop:2013_crossStripe}
\begin{equation}
E(m)/N = a_{0}+a_{1}m^{-2}+a_{2}m^{-4}\,.     \label{E_extrapo}
\end{equation}
By contrast, but as expected, other GS quantities converge less rapidly than does the energy.  For example, the magnetic order parameter, $M$, defined in the local spin coordinates by Eq.\ (\ref{gs_M}), typically follows a scheme with leading exponent $1/m$,
\begin{equation}
M(m) = b_{0}+b_{1}m^{-1}+b_{2}m^{-2}\,,   \label{M_extrapo_standard}
\end{equation}
for most systems studied to date that are either unfrustrated or
contain only moderate amounts of
frustration \cite{Kruger:2000_JJprime,Bishop:2000_XXZ,Fa:2001_trian-kagome,Darradi:2005_Shastry-Sutherland,Li:2013_chevron,Bishop:2013_crossStripe}.
Conversely, when the system is close to a QCP or when the magnetic
order parameter of the phase under study is either zero or close to
zero, the scheme of Eq.\ (\ref{M_extrapo_standard}) is usually found
to overestimate the magnetic order present and/or to yield a somewhat
too large value for the critical strength of the frustrating
interaction that is driving the corresponding phase transition.  In
such cases a scheme with leading exponent $1/m^{1/2}$,
\begin{equation}
M(m) = c_{0}+c_{1}m^{-1/2}+c_{2}m^{-3/2}\,,   \label{M_extrapo_frustrated}
\end{equation}
has then usually been found both to fit the LSUB$m$ results much
better and to yield more accurate
QCPs.\cite{DJJF:2011_honeycomb,RFB:2013_hcomb_SDVBC,Bi:2008_PRB_J1xxzJ2xxz,Bi:2008_JPCM_J1J1primeJ2,Darradi:2008_J1J2mod,Bi:2009_SqTriangle,Richter2010:J1J2mod_FM,Bishop:2010_UJack,Bishop:2010_KagomeSq,Reuther:2011_J1J2J3mod,Gotze:2011_kagome,Bishop:2012_checkerboard,Li:2012_honey_full,Li:2012_anisotropic_kagomeSq,Li:2013_chevron,Bishop:2013_crossStripe}

In general, of course, one may always test for the correct leading exponent $\nu$ for the extrapolation scheme for any GS physical quantity $Q$,
\begin{equation}
Q(m) = q_{0}+q_{1}m^{-\nu}\,,    
\label{M_extrapo_nu}
\end{equation}
by fitting an LSUB$m$ sequence of results $\{Q(m)\}$ to this form,
with each of the parameters $q_{0}$, $q_{1}$, and $\nu$ treated as
fitting
parameters \cite{Darradi:2005_Shastry-Sutherland,Bishop:2000_XXZ,RFB:2013_hcomb_SDVBC,Li:2013_chevron,Bishop:2013_crossStripe}.
For the present model we have performed fits of the form of Eq.\
(\ref{M_extrapo_nu}) for both the GS energy per spin, $E/N$, and the
magnetic order parameter, $M$, as described in Sec.\
\ref{results_sec}.  For the energy we find fitted values of $\nu$
close to 2, and thereafter use Eq.\ (\ref{E_extrapo}) to do the final
extrapolations.  Similar fits of the form of Eq.\ (\ref{M_extrapo_nu})
for $M$ have been performed before using a final extrapolation scheme
of either of the forms of Eqs.\ (\ref{M_extrapo_standard}) and
(\ref{M_extrapo_frustrated}), as appropriate. 

In Sec.\ \ref{results_sec} we first present our CCM LSUB$m$ results in
the range $0 \leq \kappa \leq 1$ for the GS quantities $E/N$ and $M$,
together with their corresponding $m \rightarrow \infty$
extrapolations, based on each of the model states N(p), N($z$), and
N-II(p) separately.  Since LSUB2 results are generally too far away
from the asymptotic limit, we perform all extrapolations with $m \geq
4$.  Furthermore, since the hexagon is such a basic structural element
of the lattice we generally prefer to do extrapolations with values of
the truncation index $m \geq 6$, whenever possible and, in particular,
when not in conflict with the clear preference that any LSUB$m$
extrapolation scheme with $n$ fitting parameters should best be fitted
with more than $n$ corresponding LSUB$m$ results.

\section{RESULTS}
\label{results_sec}
\begin{figure}
  \includegraphics[angle=270,width=8cm]{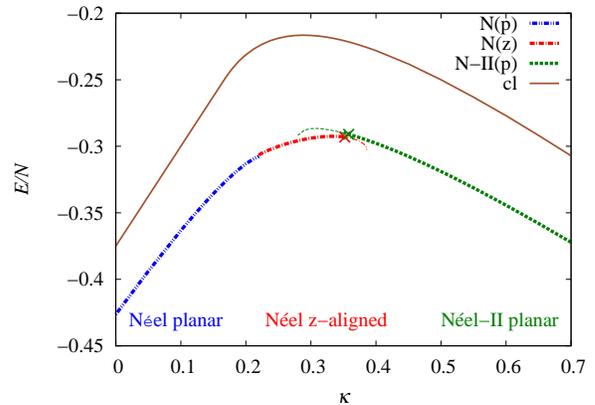}
  \caption{(Color online) Extrapolated CCM LSUB$\infty$ results for the
    GS energy per spin $E/N$ versus the frustration parameter
    $\kappa \equiv J_{2}/J_{1}$, for the spin-$\frac{1}{2}$
    $J_{1}$--$J_{2}$ isotropic $XY$ model on the honeycomb lattice (with
    $J_{1}=1$).  We show results based on the N\'{e}el planar [N(p)],
    the N\'{e}el $z$-aligned [N($z$)], and the N\'{e}el-II planar
    [N-II(p)] states as CCM model states.  The extrapolated
    LSUB$\infty$ curves shown are all based on Eq.\ (\ref{E_extrapo}),
    using LSUB$m$ results with $m=\{6,8,10,12\}$ for the N(p) and N-II(p) states, and with
    $m=\{4,6,8,10\}$ for the N($z$) state.  The times ($\times$) symbols
    mark the points where the respective extrapolations for the order
    parameter have $M \rightarrow 0$, and the unphysical portions of
    the solutions beyond those points (i.e., where $M<0$) are shown by
    thinner lines (and see text for details).  Also shown for
    comparison is the corresponding classical (cl) result from Eqs.\
    (\ref{E_np}) and (\ref{E_spiral_cl}), using the value
    $s=\frac{1}{2}$.}
\label{E}
\end{figure}

In Fig.\ \ref{E} we present our extrapolated CCM results for the GS energy per
spin, $E/N$, of the spin-$\frac{1}{2}$ Hamiltonian ${\cal H}_{XY}$ on the honeycomb lattice, where we have put $J_{1}\equiv +1$ and
$\kappa \equiv J_{2}/J_{1}$.  
Results are shown for the three separate
cases of the N\'{e}el planar [N(p)], N\'{e}el $z$-aligned [N($z$)], and
the N\'{e}el-II planar [N-II(p)] states shown in Figs.\
\ref{model}(b), (c) and (d) respectively, used as CCM model states.
For comparison purposes we also show the corresponding classical result in Fig.\ \ref{E}, taken from Eqs.\ (\ref{E_np}) and (\ref{E_spiral_cl}) with the spin quantum number set to the value $s=\frac{1}{2}$.

We do not display in Fig.\ \ref{E} the data for the individual LSUB$m$
approximations since on the scale shown the corresponding results
based on the same model state lie almost on top of each other, exactly
as was observed in our previous work for the comparable Heisenberg
model of Eq.\ (\ref{H_H}) (and see, e.g., Fig.\ \ref{E} of Ref.\
\onlinecite{RFB:2013_hcomb_SDVBC}).  Due to this rapid convergence of the
LSUB$m$ energy curves with increasing values of the truncation index
$m$, for CCM results based on the same model state, we show in Fig.\
\ref{E} only the corresponding extrapolated LSUB$\infty$ results based
on Eq.\ (\ref{E_extrapo}), which, as is usually the case, fits the
LSUB$m$ data well for each of the three model states used.

We note that for all three model states results are shown in Fig.\
\ref{E} only over certain well-defined specific ranges of the
frustration parameter, $\kappa$.  All three curves display termination
points, namely an upper one for the N(p) curve, both an upper and a
lower one for the N($z$) curve, and a lower one for the N-II(p) curve.
In turn these termination points are manifestations of the
corresponding termination points in the LSUB$m$ results, which
themselves depend on the truncation index $m$.  As is generally the
case we find here that as the index $m$ is increased the range of
values of $\kappa$ for which the respective LSUB$m$ equations (based
on a specific model state) for the CCM correlations coefficients
$\{{\cal S}_{I},{\tilde{\cal S}}_{I}\}$ have real solutions, becomes
narrower.  The termination points shown on the LSUB$\infty$ curves for
$E/N$ in Fig.\ \ref{E} are precisely those of the LSUB$m$ solution
with the highest value of the truncation index $m$ used in the
corresponding extrapolation.

Such termination points of LSUB$m$ CCM solutions are very common in
many other applications of the technique.  They have been studied in
detail and are, by now, well understood (see, e.g.,
Refs.\ \onlinecite{Bi:2009_SqTriangle,Richter2010:J1J2mod_FM,Fa:2004_QM-coll}).
They are always direct manifestations of the corresponding QCP in the
physical system being studied, at which the order associated with the
corresponding model state melts.  Hence the values $\kappa_{t}(m)$ of
the termination points for a particular end of a specific branch of
CCM LSUB$m$ solutions may, in principle, be used to estimate the
corresponding QCP for that GS phase under study, as
$\kappa_{c}=\lim_{m \rightarrow \infty} \rightarrow \kappa_{t}(m)$.  On the other hand it
is the case that the number of iterations required to solve the CCM
LSUB$m$ equations to a given accuracy increases significantly as
$\kappa \rightarrow \kappa_{t}(m)$.  Since it is correspondingly
costly, in terms of computational resources, to obtain the values
$\kappa_{t}(m)$ with the high precision necessary for accurate
extrapolations, we do not employ this method for determining the QCPs
in the present model, since we have other more accurate criteria
available to us here, as we describe below.

As is generally the case we find here too for the present $XY$ model
that the CCM LSUB$m$ results for finite values of $m$ based on the
same specified model state for each of the three states, employed here,
extend beyond the corresponding LSUB$\infty$ termination points (which
are thus the respective QCPs).  The actual LSUB$m$ termination points
for larger values of $m$ can sometimes lie very close to the
corresponding QCP at which the phase under study melts.  This is
particularly striking for the present $XY$ model for the corresponding
upper termination point for the N(p) LSUB$m$ results and the lower
termination point for the N($z$) LSUB$m$ results, both of which converge
very rapidly as $m$ is increased to what appears to be the {\it same}
value of $\kappa_{c_{1}}\approx 0.22$, as can clearly be seen from
Fig.\ \ref{E}.  It is also particularly noteworthy from Fig.\ \ref{E}
that both corresponding LSUB$\infty$ curves based on the N(p) and N($z$)
states as model states apparently meet both continuously and smoothly
at this value $\kappa=\kappa_{c_{1}}$.  Thus, based on our GS energy
results alone we have preliminary, but rather strong, evidence that
the model has a first QCP at $\kappa_{c_{1}} \approx 0.22$ from N(p) to
  N($z$) order.  The evidence from Fig.\ \ref{E} points strongly towards
  a direct $T=0$ second-order (continuous) quantum phase transition at
  $\kappa_{c_{1}}$, at which both the energy and its first derivative
  with respect to $\kappa$ appear to be continuous within very small
  error bars.

\begin{figure*}[bth]
\mbox{
\subfigure[]{\scalebox{0.31}{\includegraphics[angle=270]{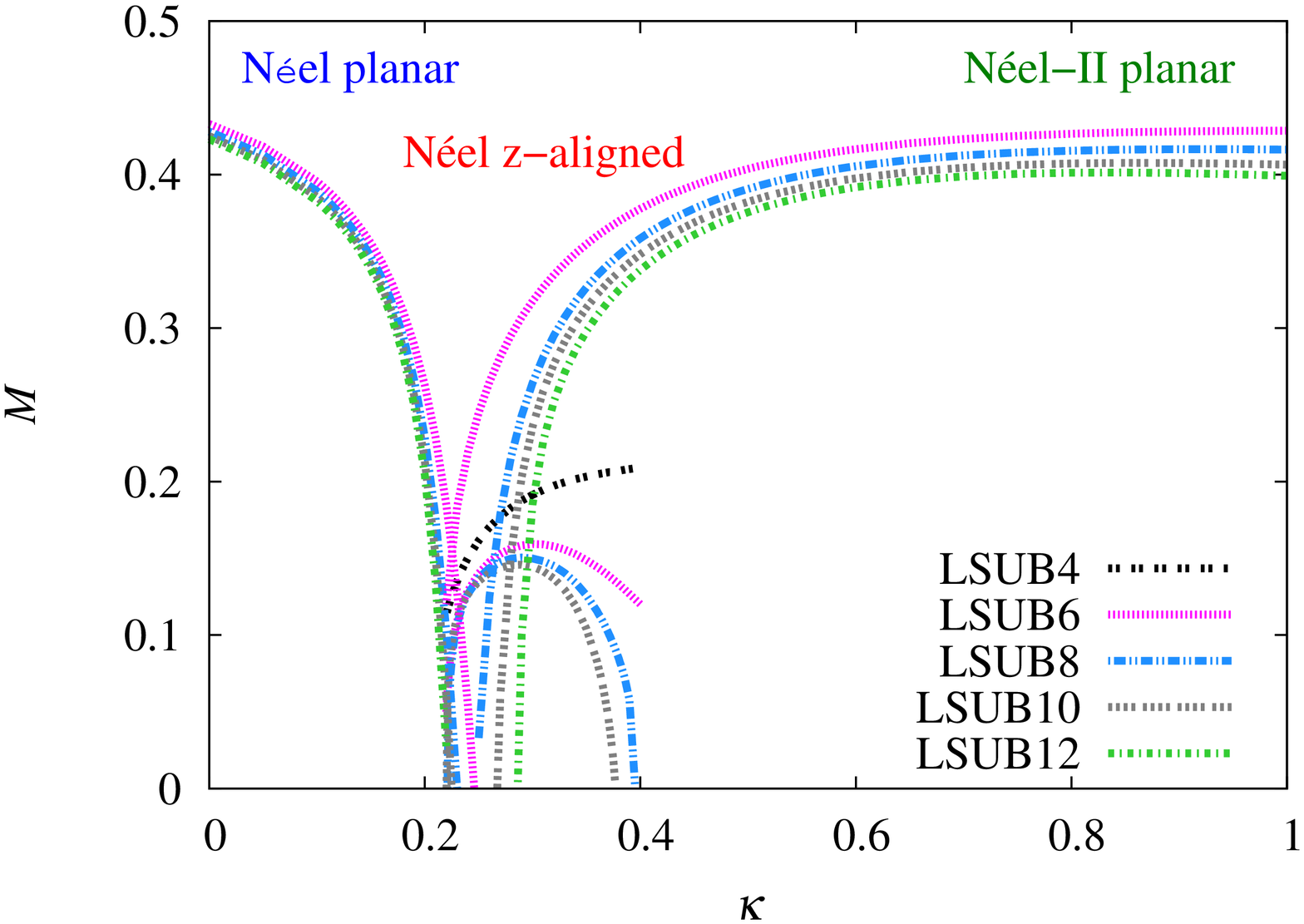}}}
\subfigure[]{\scalebox{0.31}{\includegraphics[angle=270]{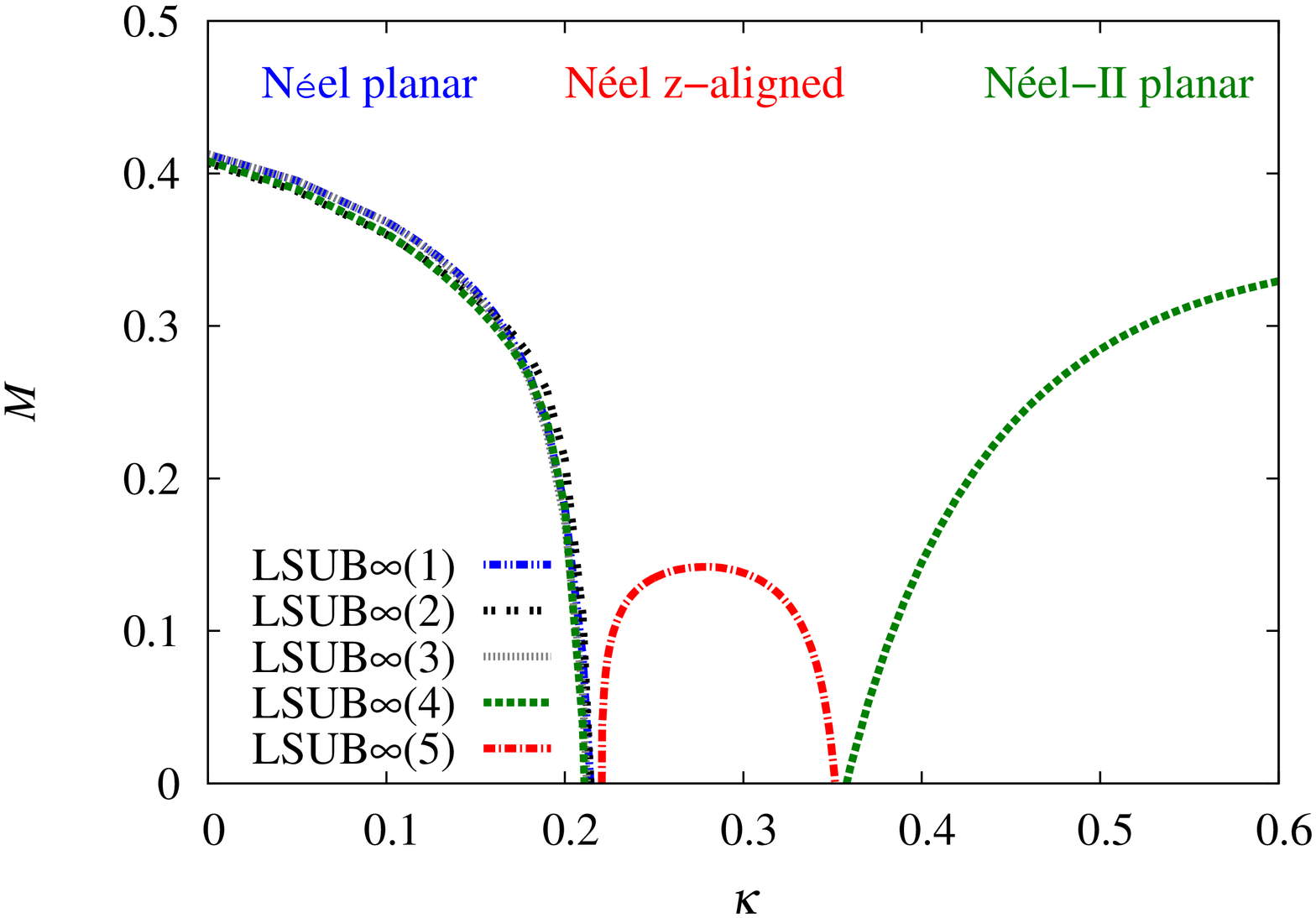}}}
}
\caption{(Color online) CCM results for the GS magnetic order $M$
  versus the frustration parameter $\kappa \equiv J_{2}/J_{1}$ for the
  spin-$\frac{1}{2}$ $J_{1}$--$J_{2}$ isotropic $XY$ model on the
  honeycomb lattice (with $J_{1}>0$).  We show results based on the
  N\'{e}el planar [N(p)], the N\'{e}el $z$-aligned [N($z$)], and the
  N\'{e}el-II planar [N-II(p)] states as CCM model states.  (a)
  LSUB$m$ results are shown with $m=\{6,8,10,12\}$ for the N(p) and
  N-II(p) states, and with $m=\{4,6,8,10\}$ for the N($z$) state.  (b)
  Extrapolated LSUB$\infty(k)$ results are shown for the N(p) state
  using both the schemes of Eq.\ (\ref{M_extrapo_standard}) with
  $k=1,3$ and Eq.\ (\ref{M_extrapo_frustrated}) with $k=2,4$, where
  the LSUB$\infty(k)$ curves with $k=1,2$ are based on LSUB$m$
  results with $m=\{6,8,10,12\}$ and those with $k=3,4$ are based on
  those with $m=\{8,10,12\}$.  For the N($z$) state the
  LSUB$\infty(5)$ curve is based on Eq.\ (\ref{M_extrapo_nu}), using
  the LSUB$m$ data set,
  $m=\{4,6,8,10\}$.  For the N-II(p) state we show only the LSUB$\infty(4)$ curve, defined as above.}
\label{M}
\end{figure*}

  In order to analyze the other termination points displayed in Fig.\
  \ref{E} (viz., the upper termination point of the N($z$) branch and
  the lower termination point of the N-II(p) branch) we first make
  some additional general remarks about the behavior of the real CCM
  solution branches near termination points $\kappa_{t}(m)$.  As has
  been observed many times previously (and see, e.g., Ref.\
  \onlinecite{Bishop:2013_crossStripe}), we also find for the present $XY$
  model that for a region near $\kappa_{t}(m)$ the respective real
  CCM LSUB$m$ solution can itself become unphysical in the sense that
  the corresponding order parameter (viz., the local average on-site
  magnetization, $M$, here) now takes negative values.  The points on
  the energy curves in Fig.\ \ref{E} where $M \rightarrow 0$, (which we
  determine as discussed below) are shown as times ($\times$) symbols, and
  the corresponding (unphysical) regions beyond those points, where
  $M<0$, are shown by thinner portions of the curves than the
  respective (physical) regions where $M>0$, which are denoted by the
  thicker portions.

  We note from Fig.\ \ref{E} that the upper critical point for the
  N($z$) phase beyond which its order parameter $M$ becomes negative is
  at $\kappa \approx 0.352$, while the corresponding lower
  critical point for the N-II(p) phase below which its order parameter
  becomes negative is at $\kappa \approx 0.358$.  However, as we
  discuss in more detail below, there is more uncertainty associated
  with the latter value.  While the two energy curves for the N($z$) and
  N-II(p) phases do not meet quite as precisely as do those for the
  N(p) and N($z$) phases, there is clear evidence from the energy
  results alone of a second QCP at $\kappa_{c_{2}} \approx 0.36$.
  Nevertheless, the energy per spin in the N-II(p) phase is still
  slightly {\it above} that of the N($z$) phase at this point, and we
  consider this difference to be outside the errors in our results.
  The simplest explanation for these results is that an
  intermediate phase exists as the stable GS phase for the system in
  the range $\kappa_{c_{2}} < \kappa < \kappa_{c_{3}}$, where
  $\kappa_{c_{3}}$ is as yet undetermined, and where the N-II(p)
  phase only becomes the lowest-energy stable GS phase (at $T=0$) for
  $\kappa > \kappa_{c_{3}}$.  We return to this point after discussing
  our corresponding CCM results for the order parameter, $M$, for the
  same three phases as shown in Fig.\ \ref{E} for the energy.

  Before doing so, however, we comment briefly on the accuracy of our
  CCM results.  For example, for the case of zero frustration
  ($\kappa=0$) with NN isotropic $XY$ interactions only, our
  extrapolated CCM value for the GS energy per spin is $E(\kappa=0)/N
  \approx -0.4263$ based on LSUB$m$ results with $m=\{6,8,10,12\}$ and
  the standard extrapolation scheme of Eq.\ (\ref{E_extrapo}).  This
  may be compared with value $E(\kappa=0)/N=-0.4261(1)$ from a
  high-order linked-cluster series expansion (SE) analysis of the
  model around the Ising limit \cite{Oitmaa:1992_honey}.  Similarly,
  at the value $\kappa=0.3$, which is around the center of the range
  where the N($z$) phase appears to be the stable GS phase, and which is
  also close to the point $\kappa=1/\sqrt{12} \approx 0.2887$
  where the classical energy is maximal, our corresponding
  extrapolated CCM result is $E(\kappa=0.3)/N \approx -0.2947$.  This
  may be compared with the recent estimated value $E(\kappa=0.3)/N =
  -0.2945(1)$ for the infinite 2D limiting case, based on large-scale
  density-matrix renormalization group (DMRG) calculations on wide
  cylinders \cite{Zhu:2013_honey_XY}.  Both of our values are in
  excellent agreement at these two points with the best results
  available by other techniques.  Exact diagonalization (ED) results
  on 24-site clusters (comprising a $4\times 2 \times 2$ torus) are
  also available \cite{Ciolo:2014_honey_XY}, yielding values
  $E(\kappa=0)/N=-0.42941$ and $E(\kappa=0.3)/N=-0.29528$, for
  example, which are also close to our values but which lie lower in
  energy than ours by about $0.7\%$ and $0.2\%$ respectively,
  presumably due to finite-size effects.

  We now turn to our corresponding CCM results for the order parameter
  $M$, and we show in Fig.\ \ref{M} the values obtained using the same
  three model states as shown in Fig.\ \ref{E} for the energy.  From
  Fig.\ \ref{M}(a) we observe that the LSUB$m$ results converge
  extremely rapidly with increasing order $m$ of approximation for the
  N(p) phase, and each of the curves tends to zero around the same
  value of $\kappa$ (near $\kappa_{c_{1}}$) with almost vertical
  slope.  The same is also true for the LSUB$m$ results near
  $\kappa_{c_{1}}$ for the N($z$) phase.  It is precisely for this
  reason that the corresponding LSUB$m$ branches of solutions based on
  both the N(p) and N($z$) model states terminate at values very close
  to the points at which the solutions become unphysical in the sense
  of yielding negative values for the order parameter, and hence why
  in Fig.\ \ref{E} the corresponding energy curves near
  $\kappa_{c_{1}}$ display no perceptible ``unphysical'' regions shown
  as thinner portions of the curves.  By contrast, the LSUB$m$ results
  for the N($z$) phase near the upper critical point converge more
  slowly than near the lower critical point, as do the overall results
  for the N-II(p) phase, especially near the (lower) critical point at
  which the N-II(p) order melts.  These latter observations thus
  provide the reason for why the N($z$) and N-II(p) curves shown in
  Fig.\ \ref{E} illustrate the ``unphysical'' regions (shown as thinner
  portions of the corresponding curves) extending beyond the points
  where the respective extrapolated values of the order parameter $M$
  have vanished.

In Fig.\ \ref{M}(b) we show extrapolated results for the order parameter of the three states used as CCM model states, based on the raw LSUB$m$ results shown in Fig.\ \ref{M}(a).
We note firstly that for the N(p) phase the extrapolated results are
extremely insensitive both to the choice of extrapolation scheme from
either Eq.\ (\ref{M_extrapo_standard}) or Eq.\
(\ref{M_extrapo_frustrated}), and to the LSUB$m$ data set used.  For
example, for the unfrustrated limiting case, $\kappa=0$, when Eq.\
(\ref{M_extrapo_standard}) is the appropriate extrapolation scheme, we
obtain extrapolated LSUB$\infty$ values $M(\kappa=0)\approx 0.4125$
when the LSUB$m$ data set $m=\{6,8,10,12\}$ is used, and
$M(\kappa=0)\approx 0.4127$ with the corresponding set
$m=\{8,10,12\}$.  These values may again be compared with, and seen to be in
excellent agreement with, the result $M(\kappa=0)=0.4133(3)$ from a high-order
SE analysis of the model \cite{Oitmaa:1992_honey}.  We have no reason
to believe that our results will be any less accurate at other values
of $\kappa$.

We observe from Fig.\ \ref{M}(b) again the extreme insensitivity of
the N(p) results to the choice of extrapolation scheme, even for the
corresponding estimates for the QCP $\kappa_{c_{1}}$ at which the
respective values for $M$ vanishes in this phase.  For example, using
Eq.\ (\ref{M_extrapo_frustrated}), which is the appropriate choice in
this regime of maximal frustration for the N(p) phase we obtain
values $\kappa_{c_{1}}\approx 0.214$ using the LSUB$m$ data set
$m=\{6,8,10,12\}$, and $\kappa_{c_{1}}\approx 0.211$ with the
corresponding set $m=\{8,10,12\}$.

By contrast with the results for the N(p) phase where we have LSUB$m$
results with $m=\{6,8,10,12\}$, for the N($z$) phase we have LSUB$m$
results only for values $m \leq 10$.  In this case it seems clearly
preferable to use the totally unbiased extrapolation scheme of Eq.\
(\ref{M_extrapo_nu}) for the order parameter $M$, where the exponent
$\nu$ is itself a fitting parameter.  The corresponding LSUB$\infty$
extrapolated result is shown in Fig.\ \ref{M}(b) for the N($z$) phase,
based on our LSUB$m$ results with $m=\{4,6,8,10\}$.  The respective
value for the lower QCP at which N($z$) order melts is thereby obtained
as $\kappa \approx 0.221$.  The same value is obtained using the set
$m=\{6,8,10\}$.  Thus, the values of the respective critical points at
which both N(p) and N($z$) forms of order melt are essentially
identical, and from the shape of the curves and, indeed, from all of
our results so far, there is very strong evidence for a lower QCP at
$\kappa_{c_{1}}=0.216(5)$ at which a second-order (continuous) phase
transition occurs between a state with N(p) order (for $0 \leq \kappa
< \kappa_{c_{1}}$) and one with N($z$) order immediately above
$\kappa_{c_{1}}$.

The GS phase with N($z$) order is then seen from Fig.\ \ref{M}(b) to
obtain a maximal value of its order parameter, $M \approx 0.142$, at
$\kappa \approx 0.279$, which is very close to the point $\kappa =
1/\sqrt{12} \approx 0.289$ where the classical version of the model has the energy taking its maximal
value.  We may also compare our own result in this phase at the point
$\kappa=0.3$, viz., $M(\kappa=0.3) \approx 0.138$, with the recent
estimate \cite{Zhu:2013_honey_XY} $M(\kappa=0.3) \sim 0.14$ for the
extrapolated 2D infinite-lattice limit, based on large-scale DMRG
calculations on wide cylinders.  Finally, for this stable N($z$) phase,
as $\kappa$ is increased further, our extrapolated LSUB$\infty$ result
shown in Fig.\ \ref{M}(b) provides a value $\kappa \approx 0.352$ for
the upper QCP at which the N($z$) order melts.

Figure \ref{M} also presents CCM results for $M$ based on the model
state with N-II(p) ordering.  Even a simple inspection by eye of the
individual LSUB$m$ results for this case, shown in Fig.\ \ref{M}(a),
shows that the LSUB6 results appear anomalous, in the sense that they
do not fit with a leading-order extrapolation scheme of the form of
Eq.\ (\ref{M_extrapo_nu}) with {\it any} value for the exponent $\nu$.
By contrast, those LSUB$m$ results with $m>6$ are accurately fitted by
such a scheme with a fitted value of $\nu$ close to 0.5, as expected
in this highly frustrated regime, over the whole range of values for
$\kappa$ shown.  Quite why the LSUB$m$ result with $m=6$ should be
anomalous for this phase is not obvious.  Very interestingly,
however, exactly the same behavior occurred for the LSUB6 results in
our prior study \cite{RFB:2013_hcomb_SDVBC} of the N\'{e}el-II phase
of the corresponding spin-$\frac{1}{2}$ $J_{1}$--$J_{2}$ Heisenberg
model of Eq.\ (\ref{H_H}) on the honeycomb lattice.  It is for these
reasons that in Fig.\ \ref{M}(b) we show for the N-II(p) phase the
extrapolated LSUB$\infty$ result for $M$ based on Eq.\
(\ref{M_extrapo_frustrated}) and on the LSUB$m$ data set
$m=\{8,10,12\}$ alone.  The corresponding estimate for the point below
which N-II(p) order vanishes is $\kappa \approx 0.358$, which is in
very close agreement with our estimate $\kappa \approx 0.352$ for the
point above which N($z$) order vanishes.  Again, all of our results
for $M$ to date indicate the presence in this model of a second QCP at
$\kappa_{c_{2}} \approx 0.355(5)$ between a N($z$)-ordered phase for
$\kappa_{c_{1}} \leq \kappa < \kappa_{c_{2}}$ and a N-II(p)-ordered
phase for $\kappa > \kappa_{c_{2}}$.

\begin{figure*}[!t]
\begin{center}
\mbox{
\raisebox{-3.7cm}{
\subfigure[]{\scalebox{0.2}{\includegraphics[]{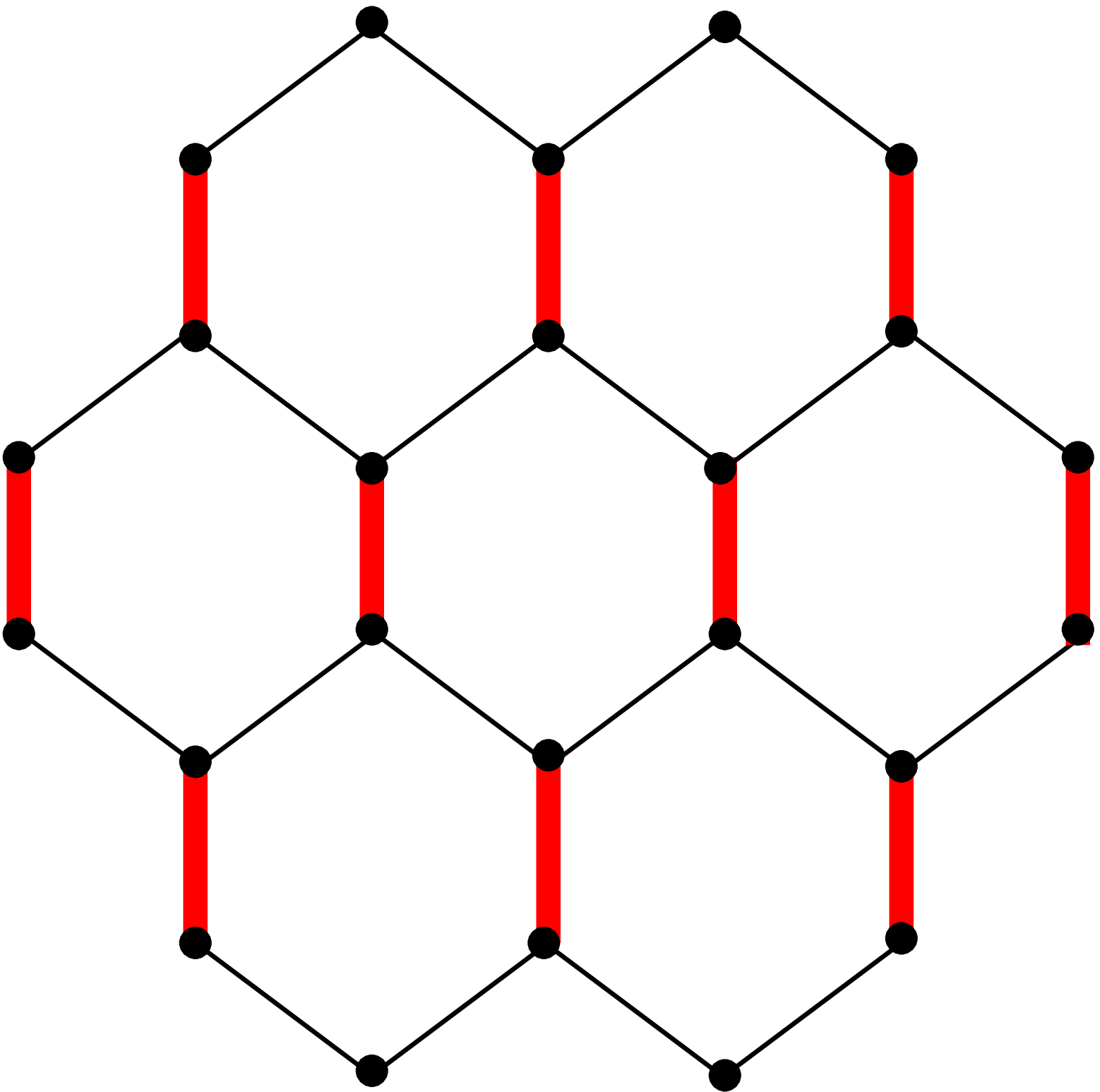}}} 
}
\subfigure[]{\scalebox{0.28}{\includegraphics[angle=270]{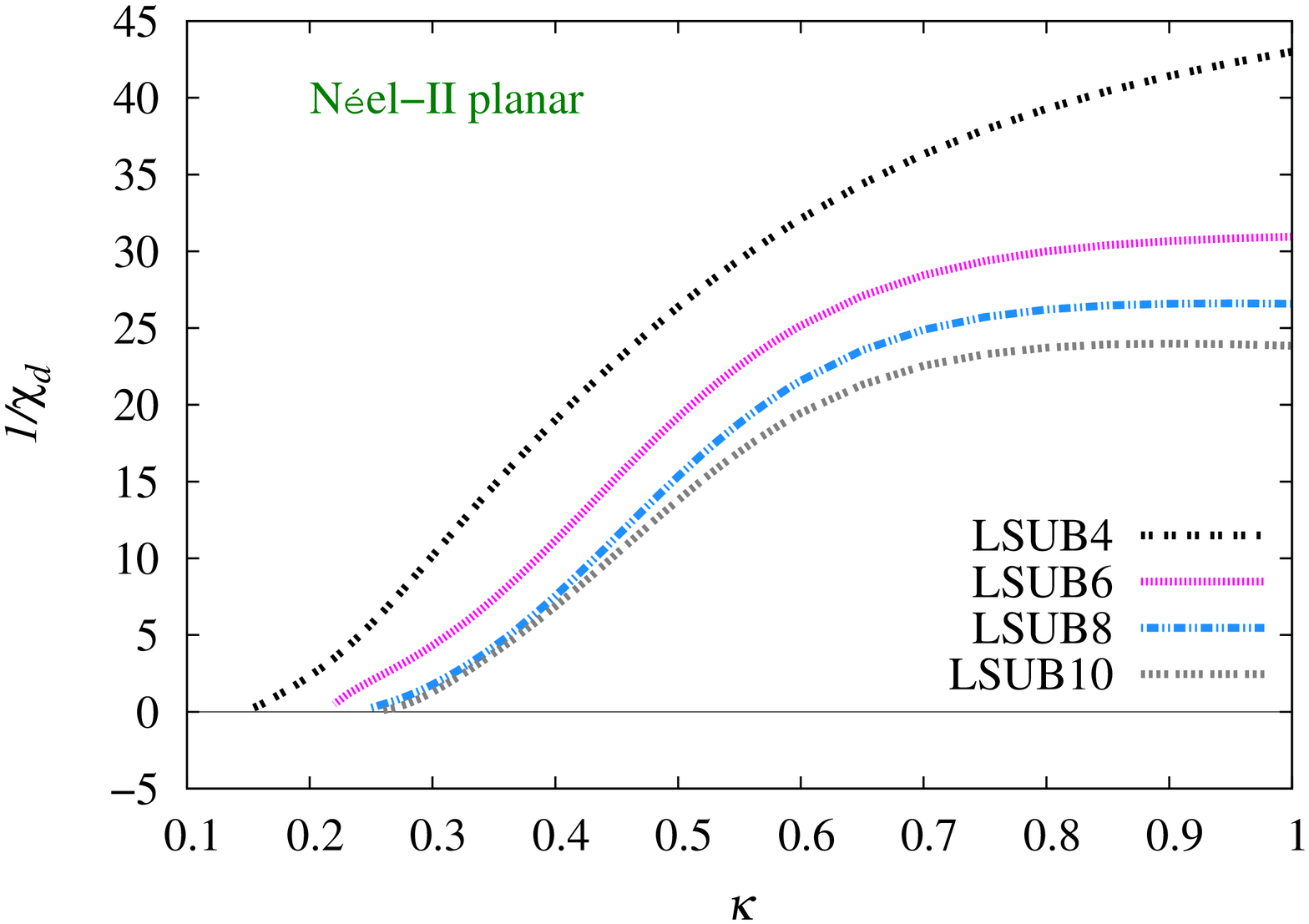}}} 
\subfigure[]{\scalebox{0.28}{\includegraphics[angle=270]{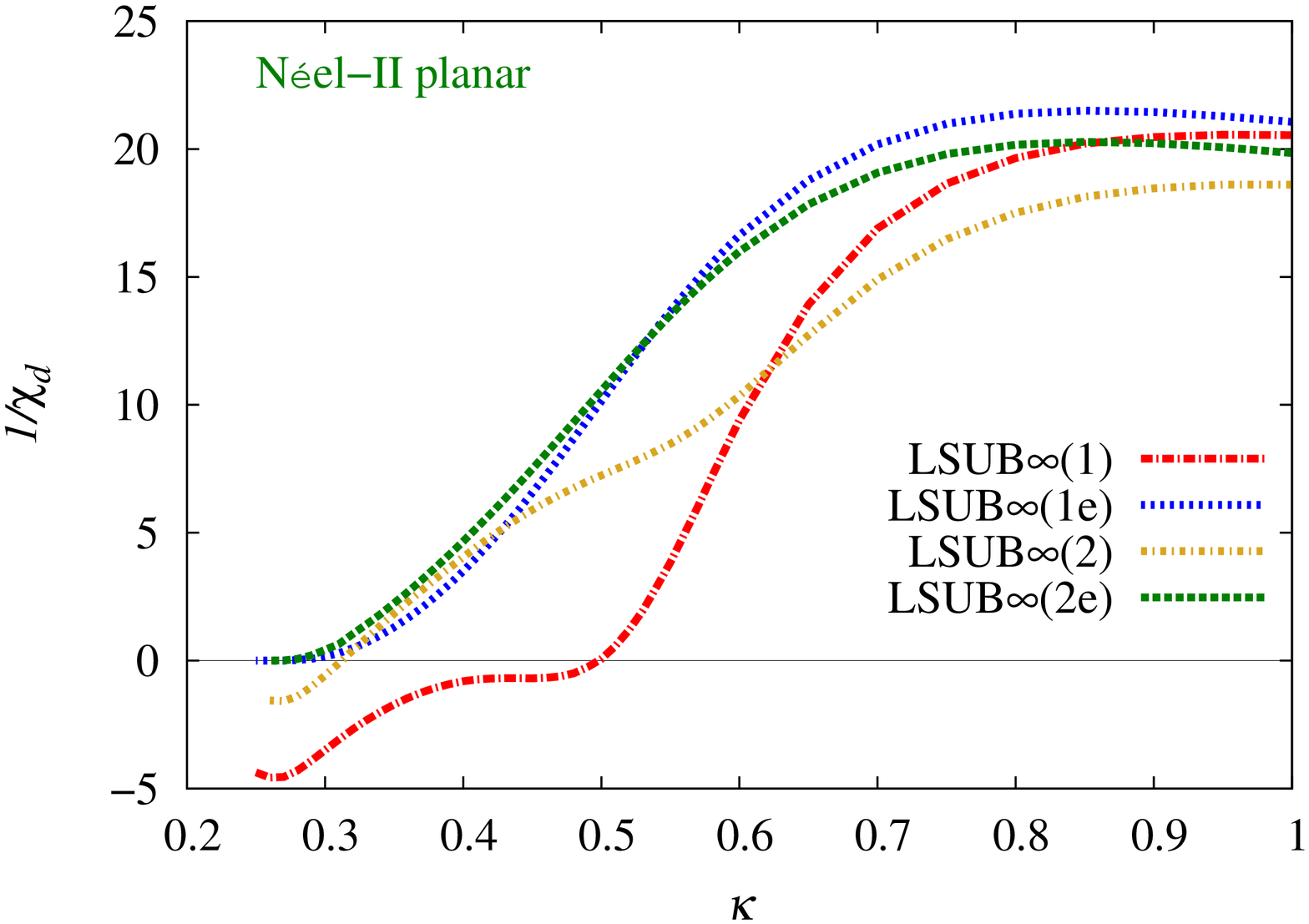}}} 
}
\caption{(Color online) (a) The perturbing field $F \rightarrow
  \delta\hat{O}_{d}\,, \delta > 0$, for the staggered dimer
  susceptibility, $\chi_{d}$.  Thick (red) and thin (black) lines
  correspond respectively to strengthened and unaltered NN exchange
  couplings, where $\hat{O}_{d} = \sum_{\langle i,j \rangle} a_{ij}
  \mathbf{s}_{i}\cdot\mathbf{s}_{j}$, and the sum on $\langle i,j
  \rangle$ runs over all NN bonds, with $a_{ij}=+1$ for thick (red)
  lines and $a_{ij}=0$ for thin (black) lines.  (b) CCM LSUB$m$
  results with $m=\{4,6,8,10\}$ for $1/\chi_{d}$ versus the
  frustration parameter $\kappa \equiv J_{2}/J_{1}$ for the
  spin-$\frac{1}{2}$ $J_{1}$--$J_{2}$ isotropic $XY$ model on the honeycomb
  lattice (with $J_{1}>0$), using the N\'{e}el-II planar state as
  model state.  (c) The corresponding extrapolated LSUB$\infty$($k$e)
  and LSUB$\infty(k)$ results, based on Eqs.\
  (\ref{Extrapo_dBonds_exponent}) and (\ref{Extrapo_inv-chi-2})
  respectively, are shown, in each case using both the LSUB$m$ data
  sets: $k=1$, $m=\{4,6,8\}$ and $k=2$, $m=\{4,6,8,10\}$.}
\label{X_SDVBC}
\end{center}
\end{figure*}    

It is interesting to note that although the LSUB$\infty$ curve for $M$
for the N($z$) phase approaches zero at $\kappa_{c_{2}}$ with a very
steep slope, indicative of a second-order transition, the
corresponding curve for the N-II(p) phase approaches zero at
$\kappa_{c_{2}}$ with a shallower slope.  On the other hand, the
individual LSUB$m$ curves for the N-II(p) phase do approach zero somewhat
more steeply.  These observations indicate that the extrapolation of
the N-II(p) results is likely to be rather sensitive in the region
around the corresponding QCP.  Since the individual LSUB$m$ curves
approach zero with increasingly steep gradients as $m$ increases, as
seen from Fig.\ \ref{M}(a), it seems entirely likely that the
LSUB$\infty$ curve should perhaps approach zero with infinite (or very
large) slope.  In this case the actual extrapolated value,
$\kappa_{c_{3}}$, of the QCP below which N-II(p) order melts would likely be
greater than the value $\kappa_{c_{2}}$ above which N($z$) order
melts.  In this case there would be an additional intermediate phase in
the regime $\kappa_{c_{2}}<\kappa < \kappa_{c_{3}}$.  Such a scenario
is entirely consistent with our previous energy results in Fig.\
\ref{E} which seem to show that at $\kappa_{c_{2}}$ the N-II(p) state
still has a slightly higher energy than the N($z$) state.

The question thus arises as to what might be the nature of such an
intermediate state.  Since real CCM LSUB$m$ solutions with finite
values of $m$ based on the N-II(p) model state clearly persist well
into any such intermediate region, one might therefore expect that the
actual GS phase in this region shares some distinct similarities with
the N-II(p) state.  One such obvious state is the so-called staggered
dimer valence-bond crystalline (SDVBC) state, also known as a lattice
nematic state, which has also been observed as a stable $T=0$ GS phase
of the analogous spin-$\frac{1}{2}$ $J_{1}$--$J_{2}$ Heisenberg model of Eq.\
(\ref{H_H}) on the honeycomb lattice \cite{RFB:2013_hcomb_SDVBC}.  Both
the N-II(p) and the SDVBC states break the lattice rotational symmetry
in exactly the same way, since the SDVBC state is basically obtained
from the N-II(p) state by replacing all of the NN parallel pairs of
spins by spin-singlet dimers, as illustrated in Fig.\ \ref{X_SDVBC}(a)

A convenient way to test for the susceptibility of a candidate GS
phase built on a particular CCM model state is to consider its
response to the imposition of a field operator $F$
(see Ref.\ \onlinecite{Darradi:2008_J1J2mod}) exactly as we did previously
\cite{RFB:2013_hcomb_SDVBC} for the corresponding case of the
spin-$\frac{1}{2}$ $J_{1}$--$J_{2}$ Heisenberg model of Eq.\
(\ref{H_H}) on the honeycomb lattice.  We thus add an extra field term
$F=\delta\; \hat{O}_{d}$ to the Hamiltonian of Eq.\ ~(\ref{H_XY}),
where the operator $\hat{O}_{d}$ now corresponds to the promotion of
SDVBC order, as illustrated schematically in Fig.~\ref{X_SDVBC}(a) and
as defined specifically in the figure caption, and where $\delta$ is
ultimately taken as an infinitesimal parameter.

The perturbed energy per site, $e(\delta) \equiv E(\delta)/N$, is
then calculated at various LSUB$m$ levels of approximation, using the
N-II(p) state as model state, for the infinitesimally shifted
Hamiltonian $H + F$.  The corresponding susceptibility of the system
to this perturbation is then defined, as usual, as
\begin{equation}
\chi_{d}
\equiv - \left. \frac{\partial^2{e(\delta)}}{\partial {\delta}^2}
\right|_{\delta=0}.
\end{equation}
The GS phase will thus become unstable against the formation of SDVBC
order when $\chi_{d} \rightarrow \infty$ or, equivalently, when
$\chi_{d}^{-1} \rightarrow 0$.  Our LSUB$m$ results for
$\chi_{d}^{-1}$ are plotted in Fig.\ \ref{X_SDVBC}(b) for
$m=\{4,6,8,10\}$, using the N-II(p) state of Fig.\ \ref{model}(d) as
the CCM model state.  We see clearly by a comparison of Figs.\
\ref{M}(a) and \ref{X_SDVBC}(b) that in each case $\chi^{-1}_{d}$
approaches zero at a value of $\kappa$ that corresponds closely to the
corresponding point at which $M \rightarrow 0$ at the same LSUB$m$
level.  It is also noteworthy that at each LSUB$m$ level, the curve
for $\chi^{-1}_{d}$ tends to zero at a rather shallow angle, and that
the intercept slope at this point decreases as the truncation
parameter $m$ increases.

Nevertheless, our LSUB$m$ results for $\chi_{d}$ (or
$\chi^{-1}_{d}$) still need to be extrapolated to the physical
(LSUB$\infty$) limit.  The simplest and most direct way to do so
\cite{Li:2013_chevron,Bishop:2013_crossStripe} is to extrapolate first
the LSUB$m$ results for the perturbed energy per spin, $e(\delta)$, using an
unbiased scheme of the form of Eq.\ (\ref{M_extrapo_nu}), namely
\begin{equation}
e^{(m)}(\delta) = e_{0}(\delta)+e_{1}(\delta)m^{-\nu}\,,   \label{Extrapo_dBonds_exponent}
\end{equation}
where each of $e_{0}(\delta)$ and $e_{1}(\delta)$, and $\nu$ are
treated as fitting parameters.  Since our standard LSUB$m$
extrapolation scheme for the GS energy is as given in Eq.\
(\ref{E_extrapo}), as discussed in Sec.\ \ref{ccm_sec}, our
expectation is that the fitted value of $\nu$ in Eq.\
(\ref{Extrapo_dBonds_exponent}) should be close to the value 2.  This has
usually been observed to be the case in previous applications, except
possibly near or inside any critical regime for the phase under
consideration, where it can deviate significantly from this value (and
see, e.g., Ref.\ \onlinecite{Bishop:2013_crossStripe} for a discussion
of a specific example of such behavior).  An alternative extrapolation
scheme
\cite{DJJF:2011_honeycomb,Li:2013_chevron,Bishop:2013_crossStripe} is
to use the LSUB$m$ results for the quantity $\chi^{-1}_{d}$ itself,
again with an unbiased scheme of the form of Eq.\
(\ref{M_extrapo_nu}), namely
\begin{equation}
\chi^{-1}_{d}(m) = y_{0}+y_{1}m^{-\nu}\,,            \label{Extrapo_inv-chi-2}
\end{equation}
with $y_{0}$, $y_{1}$, and $\nu$ treated as fitting parameters.
Corresponding extrapolations using some or all of the LSUB$m$ results
shown in Fig.\ \ref{X_SDVBC}(b) are displayed in Fig.\
\ref{X_SDVBC}(c), using each of Eqs.\
(\ref{Extrapo_dBonds_exponent}) and (\ref{Extrapo_inv-chi-2}), where
they are labelled as LSUB$\infty$($k$e) and LSUB$\infty(k)$,
respectively, and where the index $k$ denotes the LSUB$m$ data set
used in the corresponding extrapolation.

We note from Fig.\ \ref{X_SDVBC}(b) that the LSUB10 results for
$\chi^{-1}_{d}$ appears somewhat anomalous in the region $\kappa
\lesssim 0.5$ in the sense that they lie closer than expected to the
LSUB8 results by contrast with the spacings of the LSUB$m$ results
with $m \leq 8$.  Thus, clearly the LSUB10 results will not fit with a
leading-order extrapolation scheme of Eq.\ (\ref{Extrapo_inv-chi-2})
with {\it any} value of the exponent $\nu$, when used with the
corresponding LSUB$m$ results with lower values of $m$.  By contrast,
for values $\kappa \gtrsim 0.5$, the LSUB10 results fit well with the
other LSUB$m$ results in such an extrapolation.  This effect is
observed very clearly in Fig.\ \ref{X_SDVBC}(c) by a comparison of the
two curves labelled LSUB$\infty$(1) and LSUB$\infty$(2), both of which
are based on the extrapolation scheme of Eq.\
(\ref{Extrapo_inv-chi-2}), but using the LSUB$m$ results with
$m=\{4,6,8\}$ and $m=\{4,6,8,10\}$ respectively.  The LSUB$\infty(1)$
result is particularly revealing in that it produces a value of
$\chi^{-1}_{d}$ which is flat and very close to zero (actually even
slightly negative) over a range $0.36 \lesssim \kappa \lesssim 0.51$.
When we also recall that the N-II(p) phase itself becomes unphysical
(i.e., with $M<0$) in the range $\kappa \lesssim 0.36$, this result is
very suggestive indeed of SDVBC ordering becoming stable
over the regime $\kappa_{c_{2}} < \kappa < \kappa_{c_{3}}$ where
$\kappa_{c_{3}} \approx 0.51$.  Interestingly, too, the portion of the
corresponding LSUB$\infty(2)$ curve for $\chi^{-1}_{d}$ for values
$\kappa \gtrsim 0.65$ agrees reasonably well with its LSUB$\infty(1)$
counterpart and also appears to be decreasing smoothly to zero at a
similar value $\kappa_{c_{3}}$.  Over the range $\kappa_{c_{2}} <
\kappa < \kappa_{c_{3}}$ the fitted value for the exponent $\nu$ in
Eq.\ (\ref{Extrapo_inv-chi-2}) is very close to 1 for the
LSUB$\infty(1)$ curve, while it varies much more over the range $2
\gtrsim \nu \gtrsim 1$ for the LSUB$\infty(2)$ curve.  Both curves
give a value for $\nu$ which smoothly approaches the value 2 as
$\kappa$ increases beyond $\kappa_{c_{3}}$.

The LSUB$\infty$($k$e) results obtained from using the extrapolation
scheme of Eq.\ (\ref{Extrapo_dBonds_exponent}) also agree well with
the LSUB$\infty(k)$ results from using Eq.\ (\ref{Extrapo_inv-chi-2})
for values $\kappa \gtrsim 0.7$.  Both the LSUB$\infty$(1e) and
LSUB$\infty$(2e) fits yield a value $\nu \approx 2$ for the fitted
exponent for values $\kappa \gtrsim 0.5$, the value of $\nu$ then
drops sharply in both cases towards zero at the corresponding point
where $\chi^{-1}_{d} \rightarrow 0$.

It is clear that extrapolating $\chi^{-1}_{d}$ is delicate in the
range $\kappa \lesssim 0.65$.  Nevertheless, from all of our results,
taken together, it seems clear that there is a close competition
between the N-II(p) and SDVBC phases as to which provides the stable
GS phase for values $\kappa > \kappa_{c_{2}}$.  When combined with our
previous energy results, we have definite evidence that over the entire range
$\kappa_{c_{2}} < \kappa < \kappa_{c_{3}}$ the stable GS phase has
SDVBC ordering, probably mixed with N-II(p) ordering in all or part of that region; while for values $\kappa > \kappa_{c_{3}}$ the N-II(p)
state alone provides the stable GS phase.  A very precise value for
$\kappa_{c_{3}}$ is hard to predict on the basis of the present
results, but our best estimate is $\kappa_{c_{3}} \approx 0.52(3)$.

\section{SUMMARY AND DISCUSSION}
\label{summary_sec}
\begin{figure}
\begin{center}
\includegraphics[width=8cm]{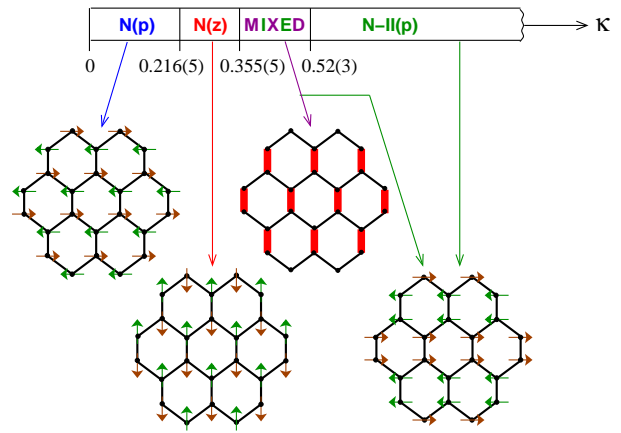}
\caption{(Color online) Phase diagram of the spin-$1/2$
  $J_{1}$--$J_{2}$ isotropic $XY$ model on the honeycomb lattice (with $J_{1}>0$ and
  $\kappa \equiv J_{2}/J_{1}>0$), as obtained by a CCM analysis.  The
  quantum critical points are at $\kappa_{c_{1}} \approx 0.216(5)$, $\kappa_{c_{2}} \approx 0.355(5)$, and $\kappa_{c_{3}} \approx 0.52(3)$, as
  shown in the diagram.  The MIXED state has SDVBC order over the whole range shown, which is probably mixed with N-II(p) ordering in all or part of the region.}
\label{phase_XY}
\end{center}
\end{figure}

In the range of values $0 \leq \kappa \leq 1$ for the frustration parameter $\kappa \equiv J_{2}/J_{1}$, we have found that the spin-$\frac{1}{2}$ $J_{1}$--$J_{2}$ isotropic $XY$ model on the honeycomb lattice has four stable GS phases at $T=0$.  They exhibit, respectively, N\'{e}el ordering in the $xy$ spin plane, N\'{e}el ordering in the $z$ spin direction, SDVBC ordering, and N\'{e}el-II order in the $xy$ spin plane, as illustrated in Fig.\ \ref{phase_XY}.  
For the corresponding isomorphic HC boson model the N(p) and N-II(p)
phases with N\'{e}el and N\'{e}el-II forms of AFM ordering in the $xy$
plane, which are both collinear spin-wave-type states, correspond to
Bose-Einstein condensates (BECs) in which the lattice bosons are
condensed, respectively, into momentum states with ${\mathbf Q} =
{\mathbf \Gamma}$ and ${\mathbf Q}={\mathbf M}^{\ast(l)};\,l=1,2,3$.  Similarly, the N($z$) state
with N\'{e}el ordering along the $z$ spin direction corresponds to a
CDW state for the HC bosons.
 
The current spin-$\frac{1}{2}$ $J_{1}$--$J_{2}$ isotropic $XY$ model
of Eq.\ (\ref{H_XY}) exhibits both similarities and distinct
differences in the structure of its GS phase diagram with that of its
spin-$\frac{1}{2}$ Heisenberg counterpart of Eq.\ (\ref{H_H}), by
contrast with the classical ($s \rightarrow \infty$) versions of the
models, which share exactly the same GS phase diagram.  In both
$s=\frac{1}{2}$ models quantum fluctuations serve to preserve the
collinear AFM N(p) order (-- but, of course, for the Heisenberg model
all directions in spin space for the N\'{e}el ordering are equally
likely --) out to larger values of $\kappa$ than for the corresponding
classical upper bound for N\'{e}el order of $\kappa_{{\rm cl}}=\frac{1}{6}$.  Our CCM estimate for this first QCP
at which N(p) order melts is $\kappa_{c_{1}}=0.216(5)$ for the present
$XY$ model, which is very close to our earlier estimate of
$\kappa_{c_{1}}=0.207(3)$ for its spin-$\frac{1}{2}$ Heisenberg
counterpart \cite{Bishop:2012_honeyJ1-J2}. For the $XY$ model our value
for $\kappa_{c_{1}}$ agrees extremely well with those of other recent
calculations, including $\kappa_{c_{1}}=0.210(8)$ from ED studies of
various 24-site clusters \cite{Varney:2011_honey_XY}, and
$\kappa_{c_{1}} \approx 0.22$ from a large-scale DMRG study on wide
cylinders \cite{Zhu:2013_honey_XY}. Both of these studies showed that
the locations of the QCPs observed were relatively insensitive to
finite-size effects.  Of course, our own CCM results pertain to the
thermodynamic ($N \rightarrow \infty$) limit, with no finite-size
scaling required.

Whereas, for the spin-$\frac{1}{2}$ Heisenberg model of Eq.\
(\ref{H_H}) on the honeycomb lattice, our CCM study
\cite{Bishop:2012_honeyJ1-J2} showed that the N\'{e}el order that
exists for $\kappa < \kappa_{c_{1}}$ gives way to PVBC order over a
range $\kappa_{c_{1}} < \kappa < \kappa_{c_{2}}$, our present CCM
study of its spin-$\frac{1}{2}$ $XY$ counterpart has shown that the
N(p) order gives way to N($z$) order.  In both models the transition
at $\kappa_{c_{1}}$ was found to be of continuous (second-order) type.
Our current CCM estimate for the second QCP at which N($z$) order
melts in the $XY$ model is
$\kappa_{c_{2}}=0.355(5)$, which may be compared with the
corresponding second QCP at $\kappa_{c_{2}}=0.385(10)$ in the
Heisenberg model at which the PVBC order melts.  Once again, our value
for $\kappa_{c_{2}}$ for the $XY$ model agrees very well with the
corresponding estimates $\kappa_{c_{2}}=0.356(9)$ from ED studies of
various 24-site clusters \cite{Varney:2011_honey_XY} and
$\kappa_{c_{2}} \approx 0.36$ from the aforementioned DMRG study
\cite{Zhu:2013_honey_XY}.

Despite the excellent agreement between the present CCM study and the
DMRG \cite{Zhu:2013_honey_XY} and ED \cite{Varney:2011_honey_XY}
studies on the values of $\kappa_{c_{1}}$ and $\kappa_{c_{2}}$, there
is disagreement over the nature of the phase between these two QCPs.
Thus, for example, whereas the DMRG study \cite{Zhu:2013_honey_XY},
which was based on much larger clusters than is possible with
available computational resources using the ED method, found clear
evidence of N($z$) ordering in this range, the ED study
\cite{Varney:2011_honey_XY} concluded that the GS phase was a QSL
state, after having specifically tested for other forms of ordering,
including of the N($z$) (or, equivalently, CDW) type, and found them
to be absent in the largest (but still small) clusters studied.  In
view of our own and the DMRG studies it does seem likely that the absence of
N($z$) order in small clusters in the region
$\kappa_{c_{1}}<\kappa<\kappa_{c_{2}}$ is a definite finite-size
effect that does not pertain in the thermodynamic limit.  Furthermore,
it is clear from Fig.\ \ref{M} that the order parameter $M$ over the
whole of the N($z$) region is rather small, and that this form of
order is relatively ``fragile''.  Indeed, as a technical aside, it is
this very fragility that makes solving the CCM LSUB$m$ equations based
on the N($z$) model state considerably more computationally
challenging, at a given level $m$ of approximation, than those based on
the other two model states.

We also note that a recent VMC study
\cite{Carrasquilla:2013_honey_XY}, which used correlated variational
QSL wave functions based on a decomposition of the underlying boson
operators into a pair of spin-$\frac{1}{2}$ fermion (parton)
operators, together with Gutzwiller projection to enforce the HC
single-occupancy constraint and a long-range Jastrow factor, found
that such QSL variational states do lie lower in energy than similar
variational AFM spin-wave states based on N(p) or N-II(p) forms of
order.  However, although such variational AFM states have been shown
to give reasonably accurate estimates for the GS energy for small
clusters by comparison with ED results (and see, e.g., Table III of
Ref.\ \onlinecite{Ciolo:2014_honey_XY}) in the N(p) regime, $\kappa <
\kappa_{c_{1}}$, for example, those based on variational QSL states in
the regime $\kappa_{c_{1}}<\kappa<\kappa_{c_{2}}$ have appreciable
inaccuracies.  Thus, for example, at $\kappa=0.3$, the extrapolated
VMC estimate for the GS energy per spin based on the QSL variational
wave function is $E(\kappa=0.3)/N=-0.28154(3)$ for the thermodynamic
limit, $N \rightarrow \infty$ (see, e.g., Table III of the
Supplemental Material of Ref.\ \onlinecite{Carrasquilla:2013_honey_XY}).
Although this energy is lower than the comparable VMC estimates found
that are based on either of the AFM [N(p) or N-II(p)] spin-wave trial
states, it is still some 4.5\% higher than our own extrapolated CCM
result based on the N($z$) model state.

It is difficult to intuit why N($z$) order should appear in the $XY$
spin model in the region $\kappa_{c_{1}}<\kappa < \kappa_{c_{2}}$, in
the complete absence of any Ising-like pairwise interactions of the
form $s^{z}_{k}s^{z}_{l}$.  In the isomorphic HC boson model this
phase manifests itself as a Mott insulator with one boson per two-site
unit cell, and the N($z$) order in the spin model translates to a CDW
order that breaks the $Z_{2}$ sublattice (inversion) symmetry of the
unit cell, in which the density $n_{{\cal A}}$ of bosons on sublattice
${\cal A}$ is higher than the density $n_{{\cal B}}=1-n_{{\cal A}}$ of
bosons on sublattice ${\cal B}$.  As we have seen, the maximum
difference occurs at a value $\kappa \approx 0.279$ of the frustration
parameter, where $n_{{\cal A}} \approx 0.642$ and $n_{{\cal B}}
\approx 0.358$.  Presumably for the HC boson model it is the HC
constraint that somehow creates the LRO in fluctuations in the density
operator in the presence of frustrated hopping, by the production of
an induced NN density-density interaction.

  Finally, just as in our earlier CCM study
  \cite{RFB:2013_hcomb_SDVBC} of the spin-$\frac{1}{2}$
  $J_{1}$--$J_{2}$ Heisenberg model of Eq.\ (\ref{H_H}) on the
  honeycomb lattice, we find here for its spin-$\frac{1}{2}$ isotropic
  $XY$ counterpart of Eq.\ (\ref{H_XY}) that for values $\kappa >
  \kappa_{c_{2}}$ there is a very close competition to form the stable
  GS phase between states with SDVBC and N-II(p) order.  A similar
  finding was also reported in the DMRG study
  \cite{Zhu:2013_honey_XY}.  While the ED study
  \cite{Varney:2011_honey_XY} found that for ($1 >$) $\kappa >
  \kappa_{c_{2}}$ the small 24-spin clusters studied exhibited N-II(p)
  ordering, explicit examination of the dimer-dimer correlation
  function $D_{ij,kl} \equiv \langle({\mathbf s}_{i}\cdot{\mathbf
    s}_{j})({\mathbf s}_{k}\cdot{\mathbf s}_{l})\rangle$ and its
  associated structure factor, showed no evidence for dimer formation
  after appropriate finite-size scaling was performed.  It seems
  clear, however, from our own results and those of others that the
  N-II(p) and SDVBC phases must lie very close in energy and these are
  unlikely to be easily resolved in calculations on finite clusters.
  Indeed in the DMRG study \cite{Zhu:2013_honey_XY}, based on much
  larger (cylindrical) clusters than are feasible by ED, the stable GS
  phase at a given value $\kappa \approx 0.5$ depended on the shape
  and size of the cylinder.  While on some cylindrical clusters the GS
  phase was the N-II(p) state, with the SDVBC state not even being
  metastable, on others the SDVBC pattern of dimer correlations was
  strongly indicated.  From all of the available evidence it seems
  that only a method such as the CCM, in which one works from the
  outset in the thermodynamic limit ($N \rightarrow \infty$), might
  have sufficient accuracy to distinguish between the comparative
  stability of the SDVBC and N-II(p) phases.  Our own best estimate is
  that the present spin-$\frac{1}{2}$ $J_{1}$--$J_{2}$ isotropic $XY$
  model on the honeycomb lattice has a third QCP at $\kappa_{c_{3}}
  \approx 0.52(3)$ such that for $\kappa_{c_{2}} < \kappa <
  \kappa_{c_{3}}$ the stable $T=0$ GS phase has SDVBC order, but which is probably mixed with N-II(p) order over all or (the higher-$\kappa$) part of this range; while for
  $\kappa > \kappa_{c_{3}}$ it has N-II(p) order, at least for the
  range $\kappa < 1$ examined.  Our earlier CCM analysis of the
  corresponding Heisenberg model of Eq.\ (\ref{H_H}) on the honeycomb
  lattice \cite{RFB:2013_hcomb_SDVBC} showed a similar QCP at a value
  $\kappa_{c_{3}} \approx 0.65(5)$.

  Within the window $0 \leq \kappa \leq 1$ we have found no evidence
  for any kinds of ordering for the present spin-$\frac{1}{2}$
  $J_{1}$--$J_{2}$ isotropic $XY$ model on the honeycomb lattice other
  than the four forms shown in Fig.\ \ref{phase_XY}.  Nevertheless, as
  $\kappa \rightarrow \infty$ the model of Eq.\ (\ref{H_XY}) reduces
  to a simple NN isotropic $XY$ model on the two independent
  triangular sublattices ${\cal A}$ and ${\cal B}$, for which the GS
  phase is known to have the 120$^{\circ}$ spin-wave ordering with
  ${\mathbf Q} = {\mathbf K}^{\ast (n)};\; n=1,2$.  Although it is
  outside the scope of this study it might still be of interest to
  apply the CCM with model states of the classical spiral type
  discussed in Sec.\ \ref{model_sec}, to test whether for any range of
  values of $\kappa$ in the region $\kappa > 1$ such quasiclassical
  spiral order is still stable under quantum fluctuations.

  In this context we note that an ED study on 24-site clusters
  \cite{Varney:2011_honey_XY} found that the 120$^{\circ}$ ordering
  was established for all values $\kappa > 1.32(2)$, whereas the
  aforementioned VMC study of the same model
  \cite{Ciolo:2014_honey_XY} found some evidence for stability of both
  spiral order (of the type found in the classical version of the model for $\kappa > \frac{1}{2}$) in the
  range $1 \lesssim \kappa \lesssim 3.5$, and 120$^{\circ}$ order in
  the entire range $\kappa \gtrsim 3.5$.  Nevertheless, it was clear
  from these investigations that the differences in energy between the
  states with spiral order and those with either N(p) or 120$^{\circ}$
  order are very small in the region where spiral order is
  variationally preferred among the class of trial wave functions
  examined.  It thus remains an interesting open question as to
  whether spiral order remains stable for the spin-$\frac{1}{2}$
  $J_{1}$--$J_{2}$ isotropic $XY$ model on the honeycomb lattice in
  the thermodynamic limit ($N \rightarrow \infty$) for any range of
  values of $\kappa$, or whether a further direct transition occurs at
  a higher QCP ($\kappa_{c_{4}} > 1$), between phases with N(p) order
  (with ${\mathbf Q}={\mathbf M}^{\ast (l)};\; l=1,2,3$) and
  120$^{\circ}$ order (with ${\mathbf Q}={\mathbf K}^{\ast (n)};\;
  n=1,2$).

  Finally, we note that in view of the differences in the $T=0$ GS
  phase diagrams of the two related spin-$\frac{1}{2}$
  $J_{1}$--$J_{2}$ models of Eqs.\ (\ref{H_H}) and (\ref{H_XY}) on the
  honeycomb lattice, it might also be of interest to investigate
  models that interpolate between them.  One simple way to do this
  would be to replace both the NN and NNN isotropic interactions by
  $XXZ$-type interactions.  In the language of the isomorphic HC boson
  model of Eq.\ (\ref{H_XY_BH}) this would be equivalent to the
  addition of off-site Ising-like two-body interaction terms proportional to
  $n_{k}n_{l}$ between NN and NNN pairs.  Such a HC
  Bose-Hubbard-Haldane type of model has been studied recently
  \cite{Varney:2012_honey_XY} by ED of clusters containing up to 30
  lattice sites, but with an effective $XXZ$ interaction only between
  NN sites (i.e., with the effective NNN interactions still of the
  isotropic $XY$ type).  An interesting related study of a spin-$\frac{1}{2}$
  $J_{1}$--$J_{2}$--$J_{3}$ model on the honeycomb lattice
  \cite{Kalz:2012_honey_anistropy}, in which third-nearest-neighbor
  exchange couplings are also included, has also been performed
  recently.  It incorporated $XXZ$ interactions on all three bonds,
  but only examined the case of ferromagnetic quantum fluctuations in
  order to avoid the minus-sign problem in the QMC simulations.
  Although, as we have indicated in Sec.\ \ref{ccm_sec}, the sign can
  be eliminated by a sublattice rotation for the $J_{1}$ and $J_{3}$
  interactions, this cannot be done for the $J_{2}$ interactions, and
  hence those results have little direct relevance to the present
  case, especially since the $J_{1}$--$J_{2}$--$J_{3}$ model was only
  studied for the special case when $J_{3}=J_{2}$.
\\
\section*{ACKNOWLEDGMENTS}
We thank the University of Minnesota Supercomputing Institute for the
grant of supercomputing facilities for this research.

\bibliographystyle{apsrev4-1}
\bibliography{bib_general}

\end{document}